\documentclass{aa}
\usepackage[varg]{txfonts}
\usepackage{graphicx}
\usepackage{multicol}
\usepackage[labelformat=simple]{caption,subcaption}
\usepackage{multirow}
\usepackage{natbib,twoopt}
\usepackage{longtable,lscape}
\usepackage{amsmath} 

\bibpunct{(}{)}{;}{a}{}{,}

\begin{document}

\title{SPOTS: The Search for Planets Orbiting Two Stars\thanks{Based on observations 
collected at the European Organisation for Astronomical Research in the Southern Hemisphere 
under ESO programmes 088.C-0291(A),  090.C-0416(A), 090.C-0416(B),  095.C-0346(A), 095.C-0346(B), 097.C-0079(A) and 097.C-0079(B) }}

\subtitle{III. Complete Sample and Statistical Analysis} 

\titlerunning{Complete Sample and Statistical Analysis}

\author{R. Asensio-Torres\inst{1}, 
        M. Janson \inst{1},
        M. Bonavita \inst{2},
        S. Desidera \inst{3},
        C. Thalmann \inst{4},
        M. Kuzuhara \inst{5, 6},
        Th. Henning \inst{7},
        F. Marzari \inst{8}
        M. R. Meyer \inst{9},
        P. Calissendorff \inst{1},
        T. Uyama \inst{10}} 

\authorrunning{R. Asensio-Torres et al.}

\institute{Department of Astronomy, Stockholm University, AlbaNova University Center, SE-106 91 Stockholm, Sweden
  \and Institute for Astronomy, The University of Edinburgh, Royal Observatory, Blackford Hill, Edinburgh, EH9 3HJ, UK
  \and INAF-Osservatorio Astronomico di Padova, Vicolo dell'Osservatorio 5, 35122 Padova, Italy
  \and Institute for Astronomy, ETH Zurich, Wolfgang-Pauli Strasse 27, 8093 Zurich, Switzerland
  \and Astrobiology Center of NINS, 2-21-1, Osawa, Mitaka, Tokyo 181-8588, Japan
  \and National Astronomical Observatory of Japan, 2-21-1, Osawa, Mitaka, Tokyo, 181-8588, Japan
  \and Max-Planck-Institut f\"{u}r Astronomie (MPIA), K\"{o}nigstuhl 17, D-69117 Heidelberg, Germany
  \and Dipartimento di Fisica, University of Padova, Via Marzolo 8, 35131 Padova, Italy
  \and Department of Astronomy, University of Michigan, 1085 S. University Avenue, Ann Arbor, MI 48109, USA
  \and Department of Astronomy, The University of Tokyo, 7-3-1, Hongo, Bunkyo-ku, Tokyo, 113-0033, Japan
}

\abstract{Binary stars constitute a large percentage of the stellar population, yet relatively little is known about the planetary systems orbiting them. Most constraints on circumbinary planets (CBPs) so far come from transit observations with the Kepler telescope, which is sensitive to close-in exoplanets but does not constrain planets on wider orbits. However, with continuous developments in high-contrast imaging techniques, this population can now be addressed through direct imaging. We present the full survey results of the Search for Planets Orbiting Two Stars (SPOTS) survey, which is the first direct imaging survey targeting CBPs. The SPOTS observational program comprises 62 tight binaries that are young and nearby, and thus suitable for direct imaging studies, with VLT/NaCo and VLT/SPHERE. Results from SPOTS include the resolved circumbinary disk around AK Sco, the discovery of a low-mass stellar companion in a triple packed system, the relative astrometry of up to 9 resolved binaries, and possible indications of non-background planetary-mass candidates around HIP 77911. We did not find any CBP within 300 AU, which implies a frequency upper limit on CBPs (1–15 $M_{\rm Jup}$) of 6–10\,\% between 30-300 AU. Coupling these observations with an archival dataset for a total of 163 stellar pairs, we find a best-fit CBP frequency of 1.9\,\% (2–15 $M_{\rm Jup}$) between 1–300 AU with a 10.5\,\% upper limit at a 95\,\% confidence level. This result is consistent with the distribution of companions around single stars.} 

\keywords{Stars: binaries -- Planetary systems -- Stars: brown dwarfs -- Stars: imaging}
\maketitle 

\section{Introduction}
Less than 1\,$\%$ of the extrasolar planets discovered to date are found to be orbiting two stars \footnote{NASA exoplanet archive, as of April 2018}. 
Given that more than half of solar-mass stars are in multiple systems \citep[e.g.,][]{janson2013, daemgen2015}, such a low number of detections might be the result of binary configurations that have been largely avoided in past planet-searching campaigns. This illustrates the technological difficulties that the most prolific methods for finding exoplanets, such as Doppler spectroscopy and transit photometry, encounter when dealing with a close binary. \par
Despite this, circumbinary planet (CBP) candidates have primarily been discovered via indirect methods, mostly with eclipsing Timing Variations (ETVs) and transits \citep{hessman2011,doyle2011,baran2015}, which has led to the first attempts to put limits on the presence of CBPs in close orbits from Kepler \citep{welsh2012} and CoRoT \citep{klag2017} data. In fact, several observational and theoretical studies point to the existence of 
a large and unexplored population of CBPs at wide separations to which the indirect methods are not sensitive. For instance, the occurrence of circumbinary protoplanetary disks where planet formation might be taking place is well demonstrated \citep[e.g.,][]{boehler2017}, and planetary systems are expected to be stable against perturbations outward of a critical radius of two to four times the binary separation \citep{holman1999}.\par
In the same way as the indirect methods, direct imaging surveys have typically excluded multistellar systems from their target list, because of the technical complications imposed on coronagraphic instruments, such as wavefront correction and 
saturation of companion stars \citep[e.g.,][]{biller2013, rameau2013, uyama2017}. However, with the right selection criteria, unresolved tight binaries observations do not suffer from deterioration in the achieved contrast, as they can essentially be treated as single stars by the wavefront sensor. In fact, close binaries might be better suited for planet searches than single stars. As the dust mass of a protoplanetary disk scales with the mass of the host 
star \citep[e.g.,][]{pascucci2016}, a pair of solar-type stars are expected to have similar amount of planet-forming material than an A-type star, while being fainter and thus providing lower detectable masses. Moreover, recent works are starting to prove that even visual binaries can be handled with the right observing strategy \citep{rodigas2015} or instrumental techniques \citep{sirbu2017}.\par
A handful of CBPs have been identified with direct imaging observations \citep[e.g., ][]{goldman2010,kuzuhara2011, currie2014, lagrange2016} on very wide orbits of hundreds or even thousands of AU.  Most of these imaged CBPs are also estimated to be massive enough to straddle the deuterium-fusing mass limit ($\sim$\,13M$\rm_{Jup}$), given the uncertainties in the computed masses \citep[for a recent summary see][]{bowler2016}. A case that exemplifies this 
situation is the circumbinary object 2M0103(AB)b, reported by \citet{delorme2013} at a distance of 84\,AU from a pair of M-type stars members of the Tucana-Horologium (THA) young moving group (YMG). This work estimated a mass of $\sim$12--14\,M$\rm_{Jup}$, but the AstraLux survey \citep{janson2017} recently re-detected the system and reassessed its age based on an updated THA age estimate by \citet{bell2015}. They predicted a mass for the circumbinary object in the range $\sim$15--20\,M$\rm_{Jup}$, above the typical planetary-mass range. In any case, the configuration of these imaged super-Jupiters impose serious challenges on planet formation theories, whether they are formed in a disk or in the fragmentation of the molecular cloud from where the host stars arose \citep{bowler2018}. \par
High-contrast imaging is thus an ideal method to search for CBPs at separations larger than the critical radius, where their orbits are expected to be stable. For this reason, the SPOTS project (Search for Planets Orbiting Two Stars;
\citep[SPOTS I]{thalmann2014}) was initiated with the goal of surveying a high number of young and nearby spectroscopic binaries with VLT/NaCo and VLT/SPHERE, and inaugurated the first demographic study of CBPs in wide orbits. \par
A description of the survey and its design was presented in SPOTS I, together with the first 27 exploratory binary observations with NaCo. The first results demonstrated the feasibility of the observing strategy
and showed several promising circumbinary candidates for follow-up observations. The second SPOTS survey paper \citep[][SPOTS II]{bonavita2016} performed a statistical analysis of CBPs around close binary stars observed
in 24 archival direct imaging surveys. In this work no substantial difference in the frequency of planetary-mass companions around tight binaries and single stars was found, and there is a substellar companion (2\,M$\rm_{Jup}$ < M$\rm_{c}$ < 70\,M$\rm_{Jup}$) best-value frequency of 6\,$\%$ within 1000 AU. These two papers also presented the complete scientific motivation for initiating the SPOTS survey. Moreover, the first discovery of the survey, the circumbinary 
protoplanetary disk around AK Sco, has recently been reported in \citet{janson2016}. The disk morphology might suggest the presence of CBPs in the system.\par
In this work of the SPOTS survey (SPOTS III), we present the full set of observations comprising 62 tight binaries, and the results of our search for CBPs. We also include a statistical analysis of the SPOTS III data alone and the combined SPOTS II + III datasets, making this work the largest direct imaging survey to date looking for planets around binary stars.

\section{Target selection}

The SPOTS target list responds not only to the science requirements but also to the technical constraints that high-contrast direct imaging observations imposes. The final target list is the result of an exhaustive literature search, followed by a careful selection aimed at excluding any wide binary as well as objects with unfavorable planet detection sensitivities. 

The fundamental criteria for target selection, previously outlined in SPOTS I, are the following: 
\begin{enumerate}
\item The binary needs to be tight enough not to affect the high-contrast observations. The basic assumption is that any pair that are not resolved by the telescope coupled to the adaptive optics system should behave as a single star. This applies to both the centering and reference process and the post-processing techniques that remove the stellar halo.\\

\item The system needs to be young enough to maximize the sensitivity to low-mass companions. This is in line with the fact that planets cool down with time \citep[e.g.,][]{marley2007}, which favors the planet-star contrast ratio at younger ages. Age estimates for the objects in the sample were based on group membership and empirical methods such as lithium abundances, chromospheric activity, or kinematics \citep{soderblom2014}. \\

\item Nearby systems are favored, since they allow us to resolve the inner circumbinary environments, that is regions as close as possible to the critical semimajor axis of stability where planetary orbits begin to be stable. Thus, neighboring binaries to the solar system can be searched for planetary-mass candidates from a distance close to the stability radius, where planets are expected to accumulate after migration \citep{welsh2014}, to very wide projected distances of hundreds of AU.\\

\item A maximum background-limited planetary detection of $\leq$\,5\,M$\rm_{Jup}$ after 1 h NaCo integration was  established as cutoff based on distance and age.
\end{enumerate}

Following this procedure, 68 tight spectroscopic binaries were selected for the entire SPOTS survey. The NaCo pilot survey imaged 28 of those targets (26 presented in SPOTS I), while 34 new ones were observed in the SPHERE-based SPOTS program.  
A total of 62 binaries were therefore observed. Follow-up observations have been acquired for a large part of the sample with candidate companions (CCs), mostly with SPHERE and over a baseline of about one year. 

 \begin{figure*}
  \resizebox{\hsize}{!}{\includegraphics{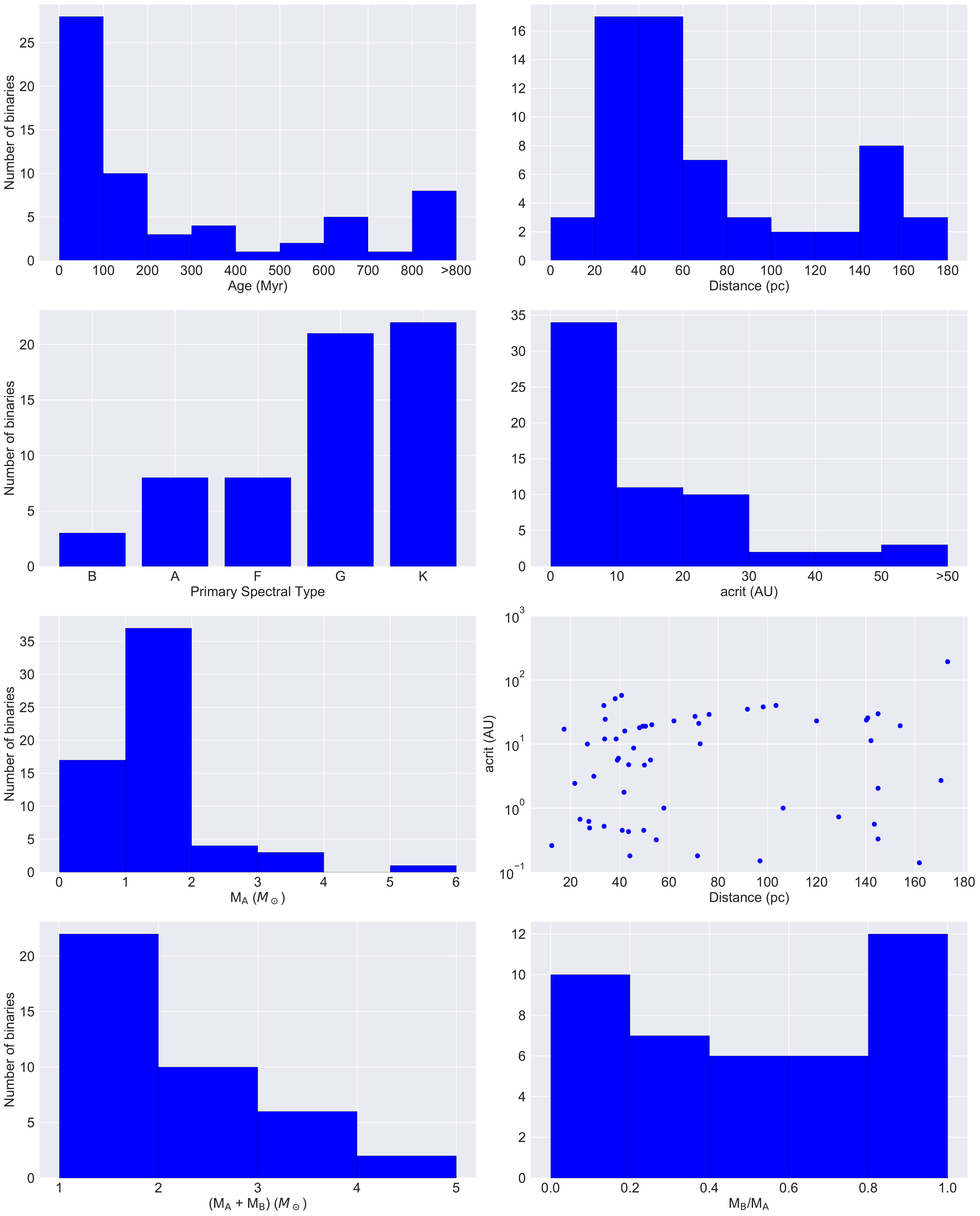}}
  \caption{Histograms representing a set of parameters that describe the selected targets for the SPOTS survey. The primary and secondary masses are taken from M$\rm_{A}$ and M$\rm_{B}$ in Table \ref{tab:mastertable}. The total mass and mass ratio is computed only for the targets for which the M$\rm_{B}$ is available.}
  \label{histogram}
\end{figure*}

\subsection{Stellar parameters}
\label{sec:stell_param}

Histograms for the age, distance and spectral type of the spectroscopic binaries chosen to be part of the survey are shown in Figure \ref{histogram}. To derive the stellar parameters we performed a literature search complemented
with the methods followed in \citet{desidera2015}. A thorough description of all the binary targets in the SPOTS survey and our choice of parameters is presented in Appendix \ref{appendix} and Table \ref{tab:mastertable}. Some of the binaries in the survey do not appear to follow the criteria for target selection previously presented, which is mainly because new observations or data have altered the initial information that we had on the targets. We however include these targets in the final survey for statistical purposes. \par
Most of the targets have estimated ages below 100\,Myr, mostly consistent with YMG or star-forming region membership. The oldest stars in our survey tend to be field stars for which no reliable age estimate can be acquired and we adopt a conservative value.  \par
When the binary is part of a stellar association, we mostly maintain the ages already adopted in SPOTS II for these groups. We included the \citet{bell2015} derived ages for the THA, $\beta$ Pic and Carina YMGs; the Columba association; and $\eta$ Cha cluster. For the Octans-Near association we took the age range from \citet{zuckerman2013} and assumed an age of 2\,Myr for the $\rho$ Ophiuchi cloud from \citet{wilking2005}. If the system does not belong to a stellar association, we estimated the age as in \citet{desidera2015} with the aid of age-dependent parameters.\par
However, tidally locked binaries may affect some of these attributes, especially with an increase in stellar rotation and chromospheric activity, which may lead to wrong age estimations. Hence, for a tidally locked system we mostly rely on lithium and kinematics indicators \citep[e.g.,][]{montes2000}. Isochronal placement is also affected by non-accounted binarity and chromospheric activity.\par
If not stated otherwise, for each derived age we approximated a corresponding mass value for the binary components using stellar evolutionary models \citep{marigo2017, baraffe2015}. In some cases we adopt the mass provided by dedicated studies in the literature that used our same age and distance estimations, and indicated this value accordingly. When dynamical masses are available, we maintain these values. For unresolved systems, however, the mass of the secondary is more difficult to estimate. In these cases, for single-lined binaries (SB1) with an available mass function, the secondary mass was derived as a minimum mass estimate for a 90\,$\deg$ inclination, and published mass ratios in double-lined binaries (SB2) were used when available. Finally, if the orbital solution is unknown and there is no information about the secondary, we assumed a mass ratio of one to compute the dynamical stability limit for CBPs.\par
Distances vary in the range $\sim$\,15--170\,pc, and have been obtained using the second GAIA data realease, Gaia-DR2, when available \citep{lindegren2018}, or otherwise with Hipparcos trigonometric parallaxes \citep{vanleeuwen2007}. For the few cases in which trigonometric parallaxes did not exist, we assumed the mean distance to the group.

\subsection{Holman-Wiegert circumbinary stability criterion}
\label{hw}

The long-term stability of planets moving in the gravitational field of a pair of stars is assessed by \citet{holman1999} who conducted a series of numerical simulations for different binary parameters and planet configurations. In the case of CBPs, they obtained a least-squares fit to the data and presented a second-order polynomial, which gives an approximation to the semimajor axis at which a hypothetical planet begins to be stable in its orbit around the central binary. This semimajor axis is referred to as the \textit{critical semimajor axis a$\rm_{c}$}, given in units of the binary separation as
\begin{align}
a_{c} =  &\, (1.60 \pm 0.04) + (5.10 \pm 0.05)\,e_{b} + ( -2.2 \pm 0.11)\,e_{b}^{2} + \notag \\
            & \,(4.12 \pm 0.09)\,\mu + ( -4.27 \pm 0.17)\,e_{b}\,\mu  + \notag \\
            & \,( -5.09 \pm 0.11)\,\mu^{2} + (4.61 \pm 0.36)\,e_{b}^{2}\,\mu^{2}  
\end{align}

where $e\rm_{b}$ is the eccentricity and \ $\mu = \frac{M_{2}}{M_{1} + M_{2}}$\  the mass ratio of the inner binary. \par
For the small number of cases in which the inner binary is surrounded by a close tertiary companion inside of which we are not sensitive to planet detection, we treat the closest pair as a single star with a mass equal to the total mass of the pair. The critical semimajor axis typically takes values between 2--4, and this is the parameter we rely on when considering 
the separation space at which one would expect to find planets around the SPOTS targets. Thus, we consider orbits interior to \textit{a$\rm_{c}$} as being unstable in the statistical analysis of the survey (see Section \ref{sec:statanaly}).\par
The computed critical semimajor axis can be seen in Figure \ref{histogram}, and individually for each target in Table \ref{tab:mastertable}. We made use of the masses as explained above, while the binary eccentricity and semimajor axis are taken from the literature, or computed
from the orbital period. If the eccentricity is not listed, we assume $e\rm_{b} = 0.5$ and, if the period is not listed and the binary has not been resolved, we take an upper limit in the binary semimajor axis as 0.1\,$\arcsec$ in projected separation (larger separations would be observable with NaCo/SPHERE). \par
It should be noted, however, that this value is only an approximation, and there will be instabilities around a$\rm_{c}$ caused by
mean-motion resonances, as already noted by \citet{holman1999}. A solution would be for instance to use the recent machine learning 
approach by \citet{lam2018}, whose neural network was able to detect these instability regions to 
$\geq$86.5\,$\%$ accuracy level. These authors however claim that regions further than 1.2 times the approximated critical 
radius a$\rm_{c}$, and closer in than 0.8, can always be classified as stable and unstable, respectively. 
We thus decided to adopt the critical semimajor axis estimation presented by \citet{holman1999}, as it is accurate enough for our statistical analysis purposes.\par
The presence of outer companions in our hierarchical systems also sets constraints on the presence of planets orbiting the space between the inner binary and the wider companions. As done for the circumbinary case, \citet{holman1999} 
developed an expression for the largest stable orbit around one star in a binary system, the so-called circumstellar critical radius $a\rm_{cs}$ in units of the binary separation, i.e.,
\begin{align}
a_{cs} =  & \, (0.464 \pm 0.006) + (-0.380 \pm 0.010)\,\mu \notag \\
            &+ ( -0.631 \pm 0.034)\,e_{b} +  \,(0.586 \pm 0.061)\,\mu\,e_{b} \notag \\
            & \,  + ( 0.150 \pm 0.041)\,e_{b}^{2}  + (-0.198 \pm 0.074)\,\mu\,e^{2.}
\end{align}

We make use of this approach to estimate the maximum distance from a binary at which a planet can be stable, given the
presence of an outer stellar companion. We thus consider the inner mass as the sum of the inner system, and the outer mass 
as the mass of the additional companion, or the sum of the individual masses if it is also a binary. Table \ref{cs} shows the
configuration of the SPOTS hierarchical systems and their computed $a\rm_{cs}$. Once $a\rm_{c}$ and $a\rm_{cs}$ are calculated, the space values considered as stable in our statistical analysis for hierarchical systems are those fullfilling \textit{a$\rm_{c}$} $<$ \textit{a$\rm_{p}$} $<$ \textit{a$\rm_{cs}$}. 
The exception is binaries orbited by a close third companion, which prevents us from resolving planets 
within the binary-companion system. In these cases we only consider orbits larger than \textit{a$\rm_{c}$}.   \\  

\begin{table*}[!htbp]
\caption{Circumstellar stability radius for the higher order systems included in the SPOTS target list.}
\label{cs}     
\centering
\begin{tabular}{l l c c c l l c}
\hline\hline
  Target ID         & M$\rm_{Target}^1$ &  M$\rm_{Outer}^2$     &  $\rho$       &  a$\rm_{CS}^3$    & Configuration     & Ref   \\
                    &   ($M_{\odot}$)   & ($M_{\odot}$)         &  (arcsecs)    &  (AU)             &                   &      \\   
\hline                      
TYC 9399 2452 1     &  0.97 + ?        & 1.69                  &   9.07        &  169.90           & Quadruple         & 1,2 \\
HIP 46637           &  0.89 + ?         & 0.87                  &  14.00        &  153.41           & Triple (wide)      & 3 \\
HIP 49669           &  3.70             & 1.10                  & 175.00        &  $>$1000          & Quadruple         & SPOTS II\\
HIP 76629           &  1.39             & 0.22                  &  10.00        &  158.65           & Triple (wide)      & 4\\
HIP 77911           &  3.14             & 0.20                  &   7.96        &  576.09           & Triple (wide)      & 5\\
HIP 78416           &  1.26 + ?          & 1.15                  &   6.55        &  217.10           & Triple (wide)      & SPOTS I \\
ScoPMS048           &  1.35 + ?          & 0.4                  &   3.38        &  152.03          & Triple (wide)      & This work, 6 \\
ROXs 43A            &  1.64             & 0.7                   &   4.48        &  256.11           & Quintuple         & This work, 7 \\
HIP 84586           &  2.05             & 0.25                  &  33.00        &  178.00           & Quadruple         & SPOTS II  \\ 
HIP 19591           &  1.27             & 0.52                  &   0.31        &    2.38               & Triple*            & This work, SPOTS I \\ 
HIP 12716           &  1.52             & 0.58                  &   0.38        &    3.16               & Triple*            & This work, SPOTS I  \\ 
HIP 7601	        &  2.03             & 1.03                  &  0.078        &     0.41              & Triple*            & 8,9      \\
HIP 105404          &  0.8             & 0.9                 &  0.04        &     0.27              & Triple*            & This work, 10\\ 

\hline
\end{tabular}
\caption*{\footnotesize $^1$ Total mass of the inner system ($M_A + M_B$ from Table~\ref{tab:mastertable}); $^2$Mass of the additional companion (or total mass of the pair in case of hierarchical systems); $^3$Outer limit for orbital stability. If no information on the mass was available (marked with a question mark), a mass equal to half of that of the primary was adopted for the computation of the dynamical stability limit. The circumstellar stability radius for the triple systems marked with a star will not be considered for the statistical analysis, as we are not sensitive to those separations. {\bf References:}  (1) \citet{koehler2001}; (2) \citet{tokovinin2010}; (3) \citet{desidera2015}; (4) \citet{nielsen2016} ; (5) \citet{kouwe2005}; (6) \citet{kraus2009}; (7) \citet{correia2006}; (8) \citet{Tokovinin2015}; (9) \citet{Tokovinin2016}; (10) \citet{guenther2005}
}
\end{table*}

\section{Observations and data reduction}
The final SPOTS survey consists of more than 90 observations of 62 binary stars, including follow-ups, over a time span of about 5.5\,years, from 2011 to 2017. The SPOTS survey has mainly relied on two high-contrast imaging instruments at the Very Large Telescope (VLT) to perform the task; these instruments are NaCo and SPHERE. The observing log is listed in Table \ref{obslog}.

\subsection{NaCo}
The initial part of the binary survey was conducted with the NAOS-CONICA (NaCo) adaptive optics facility in the $H$ band and natural guide star mode. We used the CONICA infrared camera with a 1024$\times$1024\, pixel detector (or 14\,$\arcsec$ across) and a plate scale of about 13.22\,mas per pixel. These observations explored the first 28 targets during the ESO programs 088.C-0291(A),  090.C-0416(A), and 090.C-0416(B).\par
The observation strategy and data reduction has already been comprehensively laid out in SPOTS I. As a brief summary, for each target two $H$-band four-point dither sequences of unsaturated images for photometric calibration were acquired with the neutral density filter \citep[ND\_short, 1.23\,$\pm$\, 0.05\,$\%$ transmission,][]{bonnefoy2013}, before and after the saturated scientific sequence. 
Sky frames were also observed to remove the faint background contribution in the $H$ band. Typically these backgrounds observations were noisy, in which case we constructed the background noise from the series of flux-calibration frames. The whole observing block takes about 1 hour to complete.  \par
The NaCo datasets were observed in pupil-tracking mode. In this procedure the pupil remains fixed while the sky rotates as the telescope follows the target during the observing sequence. To maximize sky rotation, the binary was preferably observed in a time window centered on its culmination. This configuration allows for angular differential imaging \citep[ADI, ][]{marois2006} post-processing recipes to remove the diffracted starlight and whiten the non-Gaussian quasi-static speckle noise pattern coming from the imperfect 
telescope and instrument optics.\par
The data reduction was fully performed with custom IDL routines. Initially, all the science frames were divided by a flat field and the sky background subtracted. The individual frames were binned to a more amenable number of images, which were centered using a moffat fit to the unsaturated wings of the stellar point spread function (PSF). Finally, PSF subtraction was applied to the stack of binned and centered frames. In this case we decided to adopt 
an aggressive version of the locally optimised combination of images \citep[LOCI,][]{lafreniere2007} algorithm with N$_{\delta}$ = 0.5\, full width at half maximum (FWHM) and 300 PSF footprints, suitable for point-like sources. We also tried the principal component analysis (PCA)-based algorithm KLIP \citep{soummer2012}, but it left uncorrected spider beam residuals that could not be properly modeled, which motivated the use of LOCI. The PSF-subtracted images were finally de-rotated to a common sky position and median-combined to obtain the starless final image. 

\subsection{SPHERE}
The rest of the binaries and almost all follow-ups were observed with the Spectro-Polarimetric High-contrast Exoplanet REsearch \citep[SPHERE,][]{beuzit2008} instrument (see Table \ref{obslog}). These observations
covered the observing periods 095.C-0346(A), 095.C-0346(B), 097.C-0079(A) and 097.C-0079(B) and went on for about two years. \par
To make these SPHERE observations fully compatible with both the NaCo-based SPOTS part in $H$ band and the SPHERE SHINE GTO survey of single stars \citep{chauvin2017}, we opted for the IRDIFS observing mode. 
This mode incorporates dual-band imaging in $H$ band with the infrared dual-band imager and spectrograph \citep[IRDIS,][]{dohlen2008}, and spectrophotometry with the near-infrared integral field spectrograph \citep[IFS,][]{claudi2008} in the Y-J range. Both of these instruments work in parallel and in pupil-stabilized ADI observations, where the star is located behind an apodized Lyot coronagraphic mask.\par
This dual-band imager of the IRDIS sub-instrument uses the H2H3 pair of narrow filters, in and out of the $H$-band methane feature ($\lambda$$\rm_{H2}$ = 1.593\,$\mu$m and $\lambda$$\rm_{H3}$ = 1.667\,$\mu$m), which is convenient for exoplanet detection.  IRDIS counts with a field of view (FoV) of $\sim$11$\times$11\,$\arcsec$ and a plate scale of 12.255\,mas and 12.250\,mas, respectively for each filter \citep{maire2016}. On the other hand, IFS has a smaller FoV (1.73\,$\arcsec$ across) and is better suited for spectral characterization purposes at low resolution $R$ $\sim$ 50. Its 3D cubes consist of 39 images of different wavelengths in the range 0.95--1.35\,$\mu$m at the same parallactic angle, i.e., with the same sky orientation. \par
For both sub-instruments, registration and flux calibration followed the same procedure. To determine the location of the star behind the coronagraphic mask, a waffle pattern was applied to the deformable mirror to produce a frame with four satellite spots forming an 'X' shape. Unsaturated images outside the coronagraphic mask were also obtained with a ND filter. These frames are used to assess the flux and nature of the found CCs.
In a first stage, we used the SPHERE Data Reduction Handling Software \citep[DRH,][]{pavlov2008} to clean the IRDIS frames from bad pixels, correct for flat field effects and subtracting the dark current. The DRH also cut out the two half-sides of the detector and centered all the H2 and H3 images using the waffle frame. For post-processing PSF-subtraction processes we adopt our own IDL-based routines, separately for each stack of H2 and H3 images. In this case both LOCI and PCA/KLIP worked well, and we used both approaches for every target. The same aggressive LOCI was used as for the NaCo data, while we kept only ten modes in the PCA algorithm. Such a small number of subtracted PCA modes is often used to search for extended structures such as disks. We checked that subtracting a bigger number of modes did not affect the achieved contrast of our SPHERE data significantly, and so we opted for these number of modes to complement the LOCI algorithm and search for both point-like structures and disks.\par
The IFS data was reduced by the SPHERE Data Center \citep{delorme2017} via the Speckle Calibration Tool (SpeCal; Galicher et al., submitted). The speckle pattern was removed with cADI \citep{marois2006}, TLOCI, \citep{marois2014} and PCA (five modes). 

\begin{figure*}
\centering
\includegraphics[width=19cm]{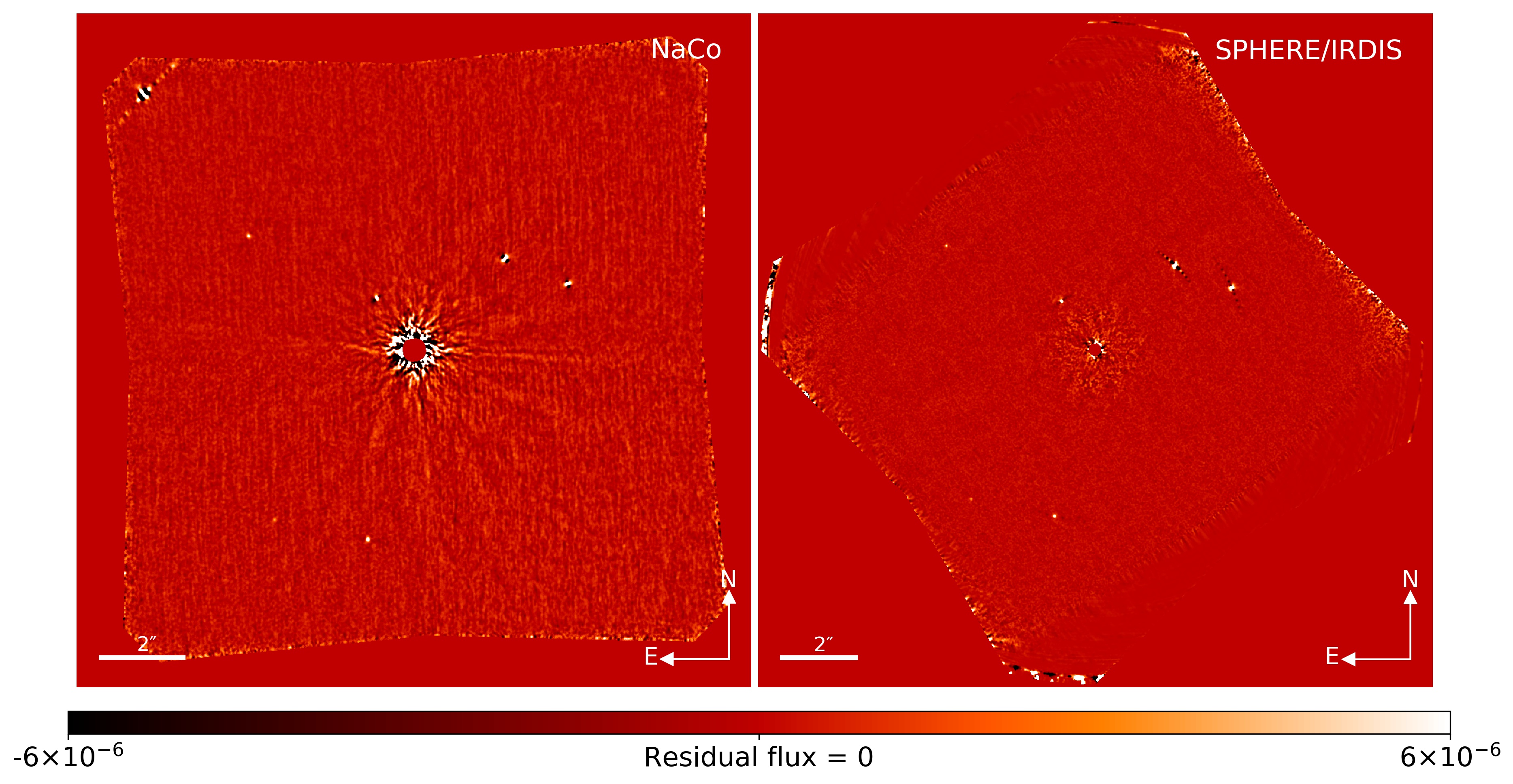}
  \caption{Example of the high-contrast final image of a SPOTS target first observed with NaCo in $H$ band (left) and later followed up with SPHERE (right, H2 filter ($\lambda$$\rm_{H2}$ = 1.593\,$\mu$m) in the dual-band IRDIS mode). The binary star is unresolved at the center of each image and its halo of diffracted light has been suppressed with the LOCI and PCA/KLIP algorithms, respectively. The figure shows the HIP 69781 system in a linear scale spanning \,$\pm$\, 6$\times$10$^{-6}$ times the primary flux. The presence of 6 CCs is revealed in both epochs, which were assigned background motion from astrometric analysis.}
  \label{im}
\end{figure*}

\subsection{Astrometry and contrast curves}
\label{astrometry}
Given the nature of the SPOTS survey where observations are spread over several years, we refrained from acquiring regular astrometric calibrations. Instead, for NaCo data we used the calibrated
true north orientation and plate scale reported in \citet[][]{chauvin2015} and \citet{schlieder2014}, whose dates are somewhat contemporary with our NaCo-based program. This approach however carries uncertainties of up to
0.4\,$\deg$ for the epochs that do not fall close in time to a reported true north value. This uncertainty corresponds to about a 4\,pixels shift from the real candidate position at distances near the edge of the NaCo FoV. This makes astrometric analysis of CCs challenging in these cases, especially for distances $>$2\,$\arcsec$, or $\sim$1\,pixel shift. For this reason, when working with NaCo data we will consider CCs that lay at projected distances of $\geq$4\,$\arcsec$ and do not match background nor common proper motion as background objects for our purposes. This fact will have to be verified with future follow-up observations. For SPHERE images we adopt the stable astrometric values derived by \citet{maire2016}, with a true north position varying by less than $\sim$0.1\,$\deg$. We also corrected for a derotator error caused by a mis-synchronization between the SPHERE and VLT clocks of $\leq$0.1--0.2 deg that affected data taken before 2016-07-03. Unlike NaCo, the SPHERE astrometric performance is accurate enough to assess the companionship for all targets within the FoV.\par
The individual position of the companions with respect to the central binary star were measured via a Gaussian fit to extract the centroid of the candidate. This procedure achieves satisfactory results in the contrast-limited regime where only Gaussian noise is expected. Sources of errors for this measurement are taken care of and explained in SPOTS I. We stress the difficulty in referencing a binary star that has been resolved or partially resolved. In a few cases the four SPHERE satellite spots did not represent well the center of the star in the science sequence. This could be caused by jitter in the differential tip-tilt-mirror in the SPHERE optical path that controls the exact location of the star. When this happened, centering was performed to the primary star based on a Moffat fit to the unsaturated wings of the stellar PSF, following the same procedure as with NaCo data. Previous uses of Moffat fitting of PSF wings in high-contrast imaging have indicated an accuracy of $\sim$\,3\,mas \citep{thalmann2011}. This is similar to the satellite spot centering accuracy of $\sim$\,2.5\,mas mentioned in the SPHERE manual. \par

Contrast curves for the binary systems in the survey are a key parameter to understand the depth and completeness of SPOTS. In 
the case of NaCo and SPHERE/IRDIS, we injected fake planets to the raw data at a S/N of 12$\sigma$, spread over the FoV of the instrument and radially separated by 2 pixels. This structure is repeated 15 times at different rotations to cover a good range of position angles. Depending on the projected distance to the star, the ADI algorithm self-subtract a different percentage of the flux of the initial injected planets. This subtraction ratio is measured at every distance and a self-subtraction curve can be created that covers the FoV. To produce the final contrast curve, the noise map of the final image, created as the standard deviation of the noise in concentric annuli around the star \citep[see, e.g.,][]{asensiotorres2016}, is divided by the unsaturated stellar flux to acquire the contrast at each distance. This preliminary contrast is finally scaled up by the self-subtraction curve for each projected distance. For SPHERE/IFS data, we made use of the 5$\sigma$ contrast provided by the $\textit{Specal}$ reduction, calculated in annuli of increasing radius with a half FWHM width.\\
The final contrast achieved is dependent on several factors. The amount of field rotation accumulated during the observation is probably the most important. If the parallactic angle variation is not big enough, the reference frame created by ADI from the stack of frames for a given image will contain a planet signal that will thus be self-subtracted, especially at close distances. The amount of time in which this field variation has been obtained also matters, as shorter periods mean more correlation and thus better subtraction of quasi-static speckles \citep{marois2006}. Target brightness, seeing conditions, and the occasional use of ND filters that block most of the photons also affect the final contrast greatly.\\

\section{Analysis and results}

\subsection{Overview}

\begin{table*}[!htbp]
\caption{Astrometry and photometry of the point sources of unknown companionship detected in the SPOTS survey. }
\label{unknowns}     
\centering
\small
\begin{tabular}{l l c c c l l l c}
\hline\hline
Target ID           & Epoch          & CC            & Sep                  &PA                  & $\Delta$$H$       & $\Delta$H2   & $\Delta$H3    & Nature \\
 	            	&	           	 &               & ($\arcsec$)          &($\deg$)            & (mag)            & (mag)        & (mag)         & if comoving\\
\hline                      
HIP 74049           & 2013-04-30    & a             & 7.31\,$\pm$\,0.04          & 221.9\,$\pm$\,0.3    & 14.09\,$\pm$\,0.14 & --                & --                       &
2\,$M\rm_{Jup}$\\
HIP 84586           &2015-04-08     & a             & 7.323\,$\pm$\, 0.007       & 109.26\,$\pm$\, 0.14   & --           & 11.87\,$\pm$\, 0.11   &  11.54\,$\pm$\, 0.11     & 5\,$M\rm_{Jup}$\\
TYC 6209-735-1      & 2015-06-08   & a             & 6.312\,$\pm$\, 0.005       & 322.59\,$\pm$\, 0.14    & --           & 9.64\,$\pm$\, 0.11   &        9.62\,$\pm$\, 0.11 & 7\,$M\rm_{Jup}$\\ 
HIP 7601            & 2015-08-25    & a             & 2.826\,$\pm$\, 0.002       & 23.10\,$\pm$\, 0.14    & --           & 7.76\,$\pm$\, 0.16    & 7.71\,$\pm$\, 0.11        & Star\\  
HIP 77911           & 2016-03-08    & a             & 4.922\,$\pm$\, 0.007       & 87.02\,$\pm$\, 0.15    & --           & 14.1\,$\pm$\, 0.9   & 14.0\,$\pm$\, 0.8           & 3\,$M\rm_{Jup}$\\
                    &               & b             & 4.230\,$\pm$\, 0.003       & 228.38\,$\pm$\, 0.14   & --          & 11.5\,$\pm$\, 1.0    &   11.3\,$\pm$\, 1.0       & 8\,$M\rm_{Jup}$ \\
                    &               & c             & 4.960\,$\pm$\, 0.005       & 233.84\,$\pm$\, 0.14    & --         & 13.2\,$\pm$\, 0.9    &   12.9\,$\pm$\, 0.8      & 4\,$M\rm_{Jup}$ \\
ScoPMS027           & 2016-03-09    & a             & 5.567\,$\pm$\, 0.007       & 338.56\,$\pm$\, 0.15   & --         & 12.0\,$\pm$\, 0.2    & 12.0\,$\pm$\, 0.2       & 4\,$M\rm_{Jup}$ \\ 
                    & 2016-04-07    & a             & 5.604\,$\pm$\, 0.005       & 338.50\,$\pm$\, 0.14    & --       & 11.6\,$\pm$\, 0.2    & 11.7\,$\pm$\, 0.2          & \\  
		            & 2016-04-07    & b             & 5.843\,$\pm$\, 0.006       & 84.85\,$\pm$\, 0.14     & --       & 11.5\,$\pm$\, 0.2    & 11.5\,$\pm$\, 0.2        & 4\,$M\rm_{Jup}$\\ 
HD 147808           & 2016-04-14    & a             & 2.215\,$\pm$\, 0.007       & 54.7\,$\pm$\, 0.2      & --       & 13.12\,$\pm$\, 0.11   & 13.17\,$\pm$\, 0.17       & 3\,$M\rm_{Jup}$\\ 
1RXS J153557.0-232417 & 2017-03-05  & a             & 3.395\,$\pm$\, 0.003       & 73.13\,$\pm$\, 0.14     & --      & 10.3\,$\pm$\, 0.3    & 10.5\,$\pm$\, 0.2            & 4\,$M\rm_{Jup}$  \\ 
				    & 2017-03-05    & b             & 4.228\,$\pm$\, 0.004       & 318.48\,$\pm$\, 0.14    & --   & 10.5\,$\pm$\, 0.2      &  10.3\,$\pm$\, 0.2           & 4\,$M\rm_{Jup}$ \\ 
	
\hline
\end{tabular}
\tablefoot{Converted to mass with COND \citep{baraffe2003} and BHAC15 models \citep{baraffe2015}
}
\end{table*}

The final SPOTS survey comprises 88 observations, including follow-ups, of 62 different binary systems, of which 
NaCo observed 19, SPHERE 34, and 9 binaries were acquired with both instruments, i.e., a first NaCo epoch was later 
followed up with SPHERE. An example of the latter case is shown in Figure \ref{im}. Two systems, CHX 18N and HIP 67199, were 
also observed as part of the survey, but later discarded as they seem to be single stars.\par
When a binary system showed a CC with a point-like morphology at a signal-to-noise (S/N) level of $\geq$\,5\,$\sigma$, this target was flagged for follow-up observations to check for common proper motion. A total of 28 binaries, that is, almost half of our sample, have been found to possess one or more CCs within the instrument FoV. We decided to consider all the CCs as potential comoving companions, except dubious CCs observed by NaCo at $\geq$4\,$\arcsec$ given its astrometric uncertainties (see Appendix \ref{appendix} for individual cases). We note that the contamination ratio scales with the square of the distance to the central binary, implying that it becomes very plausible to find false positives near the image boundary at distances $\geq$\,4--5\,$\arcsec$. Planet formation at these distances can also be put into question given the size of the protostellar disk from which the planet forms, typically not extending beyond $\sim$250\,AU \citep{andrews2007}.\par
In this way, the basic SPOTS strategy has been to obtain follow-up observations over a one year baseline at least. This was achieved successfully for the majority of the targets in the survey (see Table \ref{obslog}). For some systems that needed further observations, however, no follow-ups could be obtained during the time span of the survey. For such cases, we examined whether the detected CCs had previously been reported in the literature, and used this data when available.\par
We did not find any substellar companion among the SPOTS targets. Point sources whose companionship could not be tested are listed in Table \ref{unknowns}. All the conditional planetary-mass candidates 
are found at large distances ($>$\,2\,$\arcsec$), which makes it rather unlikely that these objects are comoving with the central binary. For a complete description of all the 
targets in the survey and the individual results of our observations, see Appendix \ref{appendix}.\par
\begin{figure}
  \resizebox{\hsize}{!}{\includegraphics{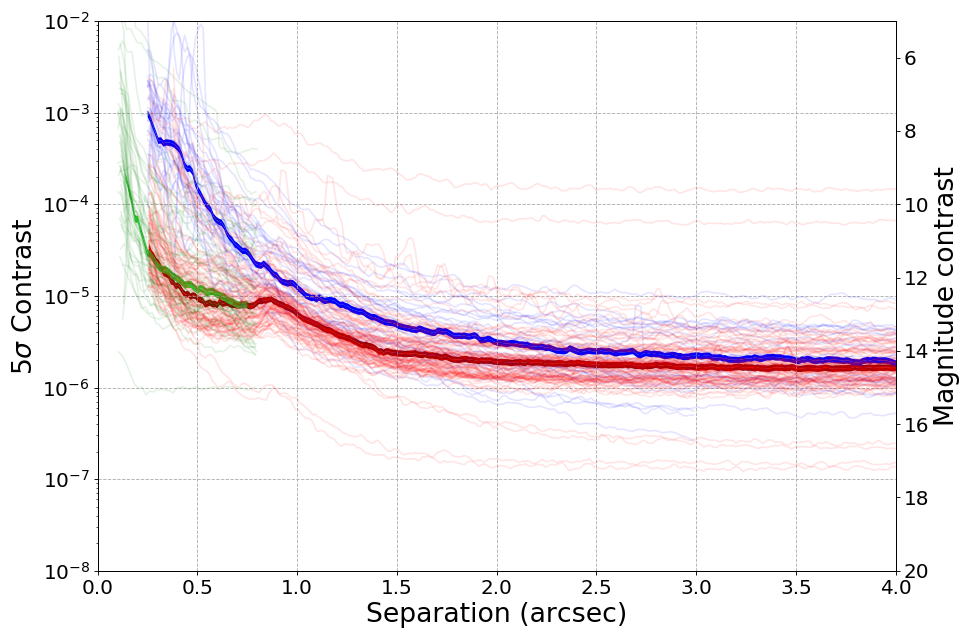}}
  \caption{Contrast curves for the targets in the SPOTS survey. Blue, red and green light curves are the individual contrast curves for the binaries observed by NaCo (LOCI), SPHERE/IRDIS (best of LOCI and PCA 10 modes) and SPHERE/IFS (best of TLOCI and PCA 5 modes), respectively. The thick curves correspond to their median
  values. }
  \label{contrast}
\end{figure}

An important auxiliary science product is the capacity of our survey to detect the scattered light emission coming off circumbinary disks grains. The SPOTS survey resolved the protoplanetary disk around AK Sco between $\sim$13--40\,AU (see \citet{janson2016} for a thorough description of the system) with SPHERE, whose morphology may be explained by an eccentric ring or the existence of two spiral arms. These two interpretations might imply the presence of unseen CBPs. \\
The 5$\sigma$ contrast curves for all the SPOTS targets are shown in Figure \ref{contrast}, which are essential to probe the mass-distance detectability region of the circumbinary companions in the survey. As expected, SPHERE performs significantly better, about a factor of 20, than NaCo in the speckle-dominated inner region. Median magnitude contrasts go down from $\sim$7.5 and $\sim$11 at 0.25\,$\arcsec$, for NaCo and SPHERE respectively, to $\sim$14.5 at distances $>$3\,$\arcsec$ for both instruments. The SPHERE/IFS configuration attains a better IWA, reaching contrasts of $\sim$10$^{-4}$ at $\sim$0.1\,$\arcsec$, although it does not improve the contrast achieved by IRDIS in the overlapping region. Using SDI might improve the contrast in this range, but the results would be less straightforward to interpret since the self-subtraction would depend on the unknown flux distributions of potential companions in the image. \par
This contrast curve analysis for such a large amount of targets gives us a meaningful evaluation of the performance of these two instruments, especially for a last-generation imager such as SPHERE. The NaCo brightness ratio agrees well with previous NaCo-based surveys of nearby stars \citep[e.g.,][]{chauvin2015}, and it is comparable to other first-generation dedicated high-contrast instruments such as Subaru/HiCIAO-AO188 \citep{janson2013seeds,uyama2017} and Gemini/NICI \citep{biller2013}. This result demonstrates the successful SPOTS observation strategy, showing that, with the right target selection, the current high-contrast imaging capacity is not harmed by binarity.  \\

\subsection{Resolved binaries}
The astrometry on the resolved SPOTS binary systems are presented in Table \ref{binaries}. We use the NaCo and SPHERE unsaturated frames to measure the stellar companion-primary relative position. 
We also include wider stellar circumbinary companions orbiting unresolved inner binaries (UX For, ROXs 43A, ScoPMS048 and TYC 8104 0991 1). Three of these systems, Alhena, UX For, and V1136 Tau, were already presented in SPOTS I. Our results
differ slightly from those presented there, probably because a mismatch in the sign of the true north NaCo value. To the best of our knowledge, the $\tau$ Hya and TYC 8104 0991 1 systems are resolved for the first time; the SBI $\tau$ Hya show a $\sim$10\,$\deg$ orbital motion over a baseline of about a year.\\

\begin{table*}[!htbp]
\caption{Relative astrometry of the spatially resolved binaries in the SPOTS survey (including close triple systems). First detection in boldface. }
\label{binaries}     
\centering
\begin{tabular}{ l l l l l l l }
\hline\hline
Target ID & Epoch &  Sep     &PA   &  $\Delta$$H$   & $\Delta$$H2$   & $\Delta$$H3$ \\    
 		&           & (mas)   &($\deg$) & (mag)    &  (mag)       & (mag)       \\
\hline        
HIP 16853   & 2011-11-09    &  2689\,$\pm$\,4      & 89.32\,$\pm$\,0.05     & 4.17\,$\pm$\,0.13 & -- & -- \\
HIP 31681   & 2011-12-22 & 382.1\,$\pm$\,1.1 & 260.75\,$\pm$\,0.14     & 3.36\,$\pm$\,0.07 & -- & --  \\ 
	        & 2013-03-01 & 379\,$\pm$\,4 & 258.7\,$\pm$\,0.5     & 3.26\,$\pm$\,0.13 & -- & --  \\
HIP 19591   & 2012-11-21 & 248\,$\pm$\,2 & 199.1\,$\pm$\,0.4     & 0.97\,$\pm$\,0.04 & -- & -- \\ 
HIP 12716   & 2013-01-05 & 305\,$\pm$\,3 & 166.2\,$\pm$\,0.4     & 2.29\,$\pm$\,0.10 & -- & --\\ 
\textbf{TYC 8104 0991 1} & 2013 -01-07 & 133\,$\pm$\,3 & 221.7 \,$\pm$\,  1.2 & 1.14\,$\pm$\,0.05 & -- & --\\
HIP 7601	& 2015-08-25 &  88.08\,$\pm$\,0.18 & 286.92\,$\pm$\,0.18     & -- & 0.68\,$\pm$\,0.05 & 0.57\,$\pm$\,0.09  \\
HIP 12225   & 2015-08-27 &  73.30\,$\pm$\,0.11 &  65.43\,$\pm$\,0.16     &  -- & 0.36\,$\pm$\,0.16 &  0.34\,$\pm$\,0.17 \\ 
\textbf{HIP 46509}  & 2015-12-20  & 342\,$\pm$\,2 & 96.7\,$\pm$\,0.4 &  -- & 4.7\,$\pm$\,0.9 & 4.6\,$\pm$\,0.8 \\ 
			& 2017-01-11 & 395.9\,$\pm$\,1.7 &  86.1\,$\pm$\,0.3     & --    & 4.68\,$\pm$\,0.05 & 4.57\,$\pm$\,0.10 \\
ScoPMS048   & 2016-04-03 &3389\,$\pm$\,3 & 191.82\,$\pm$\,0.14     & -- & 2.37\,$\pm$\,0.03 & 2.26\,$\pm$\,0.03 \\
ROXs 43A     & 2016-08-27 & 295.3\,$\pm$\,1.3 & 156.8\,$\pm$\,0.3     &  -- & 3.50\,$\pm$\,0.05 & 3.32\,$\pm$\,0.04 \\
\hline
\end{tabular}
\end{table*}

\subsection{Case of HIP 77911}
\label{sec:77911}
\begin{figure}
  \resizebox{\hsize}{!}{\includegraphics{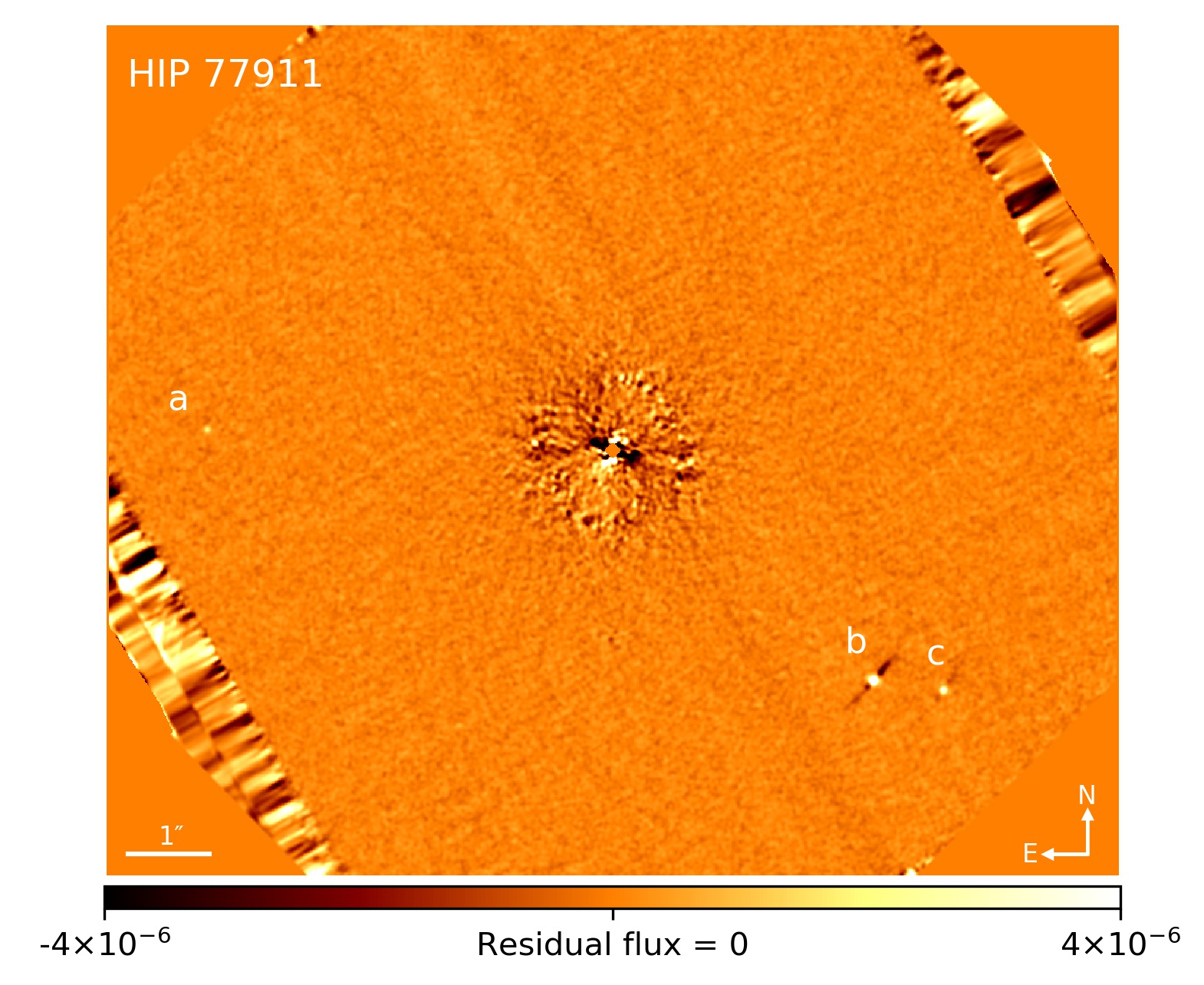}}
  \caption{HIP 77911 system as observed with SPHERE/IRDIS with the $H2$ filter. Up to 3 CCs can be seen in the image; these have estimated masses between 3 and 8\,$M\rm_{Jup}$ if they were comoving with the central binary. This image has been reduced with PCA. }
  \label{hip77911}
\end{figure}

\begin{figure*}
  \resizebox{\hsize}{!}{\includegraphics{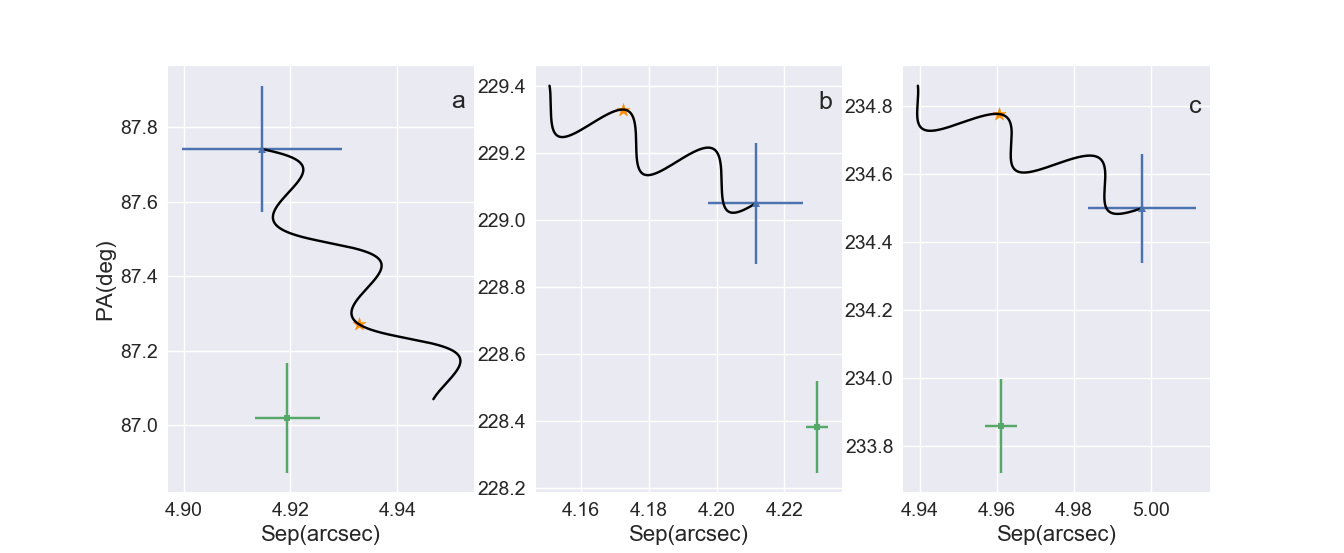}}
  \caption{Astrometric study of the HIP 77911 CBPs candidates. Each panel shows the astrometry of the \textit{a}, \textit{b}, and \textit{c} candidates, respectively, indicated in the top right corner. The y-axis shows the position angle with respect to the central binary in degrees, and the x-axis indicates separation in arcseconds. The position of the targets in the HiCIAO epoch is shown as a blue triangle, and the black curve emerging from it represents the position where a background
  point source would be found with time. The orange star points out the position of the candidate at the later SPHERE epoch if it was moving with the background. The green square is the position at which it was actually found.}
  \label{astro_hip77911}
\end{figure*}

This close binary in the Upper Scorpious (US) subregion includes a B-type primary and a wide low-mass stellar companion at about 8\,\arcsec ($\sim$1300\,AU in projected separation). Our only observation of this system
was conducted with SPHERE in March 2016. Although we only collected 5.6\,deg of sky rotation, our reduction revealed the presence of three planetary-mass CCs between 4 and 5\,$\arcsec$ (see Figure \ref{hip77911} and Table \ref{unknowns}). We note that this is in principle outside the stability space area given by the presence of the inner binary and outer stellar companion, as the circumstellar stability radius would be located at about 576\,AU or 3.56\,$\arcsec$; see Table \ref{cs}. However, we only have a lower limit for the mass of the secondary, and have no information about the eccentricity or physical separation of the orbit of the wide companion around the binary star. We recall that the stability criterion is merely statistical, and cannot be treated as a certain value. \par
HIP 77911 was also previously observed in the Subaru Strategic Exploration of Exoplanets and Disks (SEEDS) project, as reported in \citet{uyama2017}. These ADI images of the HIP 77911 system were obtained with HiCIAO \citep{Suzuki_2010_HiCIAO}. 
To check for common proper motion, we compare the astrometry of the three CCs around HIP 77911 in the SPHERE and HiCIAO images. Indeed, \citet{uyama2017} identified some point sources at separations greater than 4\arcsec, although they assumed these were background sources because of their large separation, and thus did not report the astrometric values.\par
We therefore rechecked the HIP 77911 images observed with HiCIAO.  Since the very small proper motion of HIP 77911 ($\Delta$\,RA = -13.6\,mas/yr, $\Delta$\,Dec = -22.7\,mas/yr) needs an accurate astrometry, we attempted to update the HiCIAO ADI data reduction of the HIP 77911 dataset observed in June 2014, using the corrections for the optical distortion, plate scale, and angle of the true north orientation. As described in \citet{Brandt_2013_ACORNS} and \citet{Helminiak_2016_V450AND}, the corrections were derived by comparing the HiCIAO observation of the M5 globular cluster, which was observed in the same run as HIP 77911, with the archival M5 image of ACS on {\it Hubble Space Telescope}.  With the well-calibrated astrometry corrections, the HiCIAO data of HIP 77911 were reduced using the ACORNS pipeline \citep{Brandt_2013_ACORNS}. We then applied the full data reductions in ACORNS, including the LOCI PSF subtraction, to the HiCIAO data of HIP 77911.\par
We confirmed that the CCs detected in the SPHERE image exactly appear in the HiCIAO image, and measured their separation and position angles relative to the parent star, which are shown in Figure \ref{astro_hip77911}. Only candidate \textit{a} appears to be consistent with background motion within error bars. Candidate \textit{b} moves outward, and is closer to common proper motion. Candidate \textit{c}, on the other hand, moves inward toward the inner binary as a background object would, but its position angle is far from the background trajectory. Given that it is difficult to assess the astrometry of a system observed with two different instruments, we tried to use the wide stellar companion seen in the HiCIAO image to calibrate its plate scale and position angle, assuming that it has not moved in its orbit since the value reported by \citet{kouwe2005} with ADONIS in June 2001.
Indeed, assuming a semimajor axis of 1300\,AU, the stellar companion would move at a rate of 0.2\,AU/yr, corresponding to 2.6\,AU or a maximum (if radial) motion of $\sim$16\,mas in projected separation over the $\sim$13\,yrs that went by between the ADONIS and HiCIAO observations. That is almost three times less than the motion expected from the proper
motion of the HIP 77911 system between the HiCIAO and SPHERE epochs ($\sim$44\,mas).\par
However, the ADONIS astrometry is not accurate enough for our purposes, with uncertainties in the plate scale of about $\sim$50\,mas at the position of the tertiary companion. We might think that HiCIAO's plate scale is slightly off, and speculate with a slightly different value, but none of the possibilities correspond with the three CCs being comoving or background stars, as seen in Figure \ref{astro_hip77911}.\par
With this situation, we cannot confidently state what the nature of these CCs is. The system however hints the existence of a potential non-background behavior unrelated to astrometric calibrations. The fact that candidates \textit{b} and \textit{c} behave in such a different way is especially puzzling because they are so close to each other in the sky. We then advise a future follow-up of HIP 77911 to shed light on these three point sources. The position of the wide stellar companion will certainly be helpful to calibrate the astrometry of HiCIAO if observed with a third instrument with a FoV wide enough to include it.

\subsection{New low-mass stellar companion to the $\lambda$ Muscae binary system}
\label{lambdamus}

\begin{figure*}
\centering
\includegraphics[width=18cm]{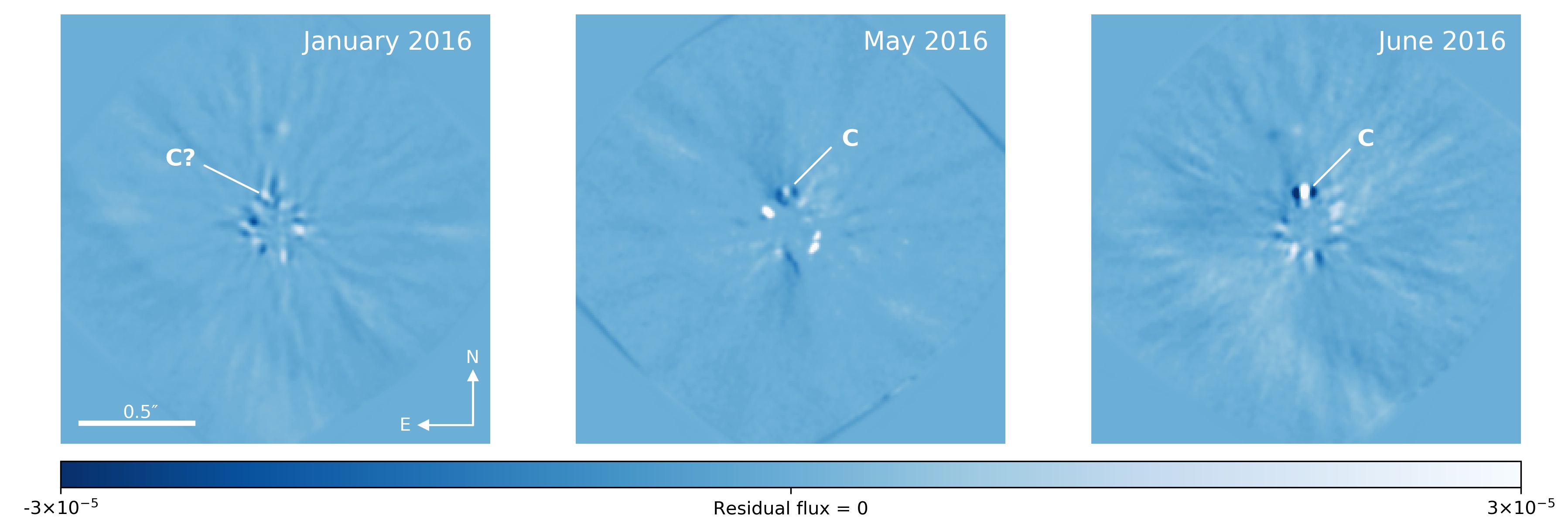}
  \caption{SPHERE/IFS YJ image of the $\lambda$ Muscae system revealing the presence of a very close companion. The three epochs for which SPOTS observed this target are shown in a linear stretch spanning \,$\pm$\, 3$\times$10$^{-5}$ times the primary flux. The unresolved binary is located at the center of each frame and the PSF halo is removed with cADI. The found comoving companion is indicated with a "C", except for January where the detection is ambiguous.}
  \label{triple}
\end{figure*}

We report the SPOTS discovery of a low-mass stellar companion at a very close angular separation to the astrometric $\lambda$ Muscae binary, a 740\,Myr old system formed by an A-type primary SB1 in a 1.6-year orbit. The system was observed with SPHERE during three different epochs in January, May, and June 2016. The first epoch did not collect enough field rotation to detect the companion at >\,5$\sigma$, but it was unveiled in both IRDIS and IFS modes in the two latter observations. The weather conditions in the May dataset were poor with a seeing above 1\,$\arcsec$, which was the reason why the observation was repeated in June.\par
Figure \ref{triple} shows the IFS final images for the three epochs and the detection of a companion at a projected separation of only $\sim$155\,mas or $\sim$6\,AU. This is indeed a new companion, as the relative semimajor axis of the inner binary is 1.54\,AU (39.5\,mas), with a separation at apoastron of 52\,mas, which is about three times smaller than the distance at which we observe the companion. Although this object is a stellar companion, it is representative to see that this separation corresponds to just outside the critical semimajor axis for stability derived for this system if we only consider the inner binary; this is in line with Kepler's discoveries with observed planets laying just outside the critical radius boundary \citep{welsh2014}.\par
\begin{figure}
  \resizebox{\hsize}{!}{\includegraphics{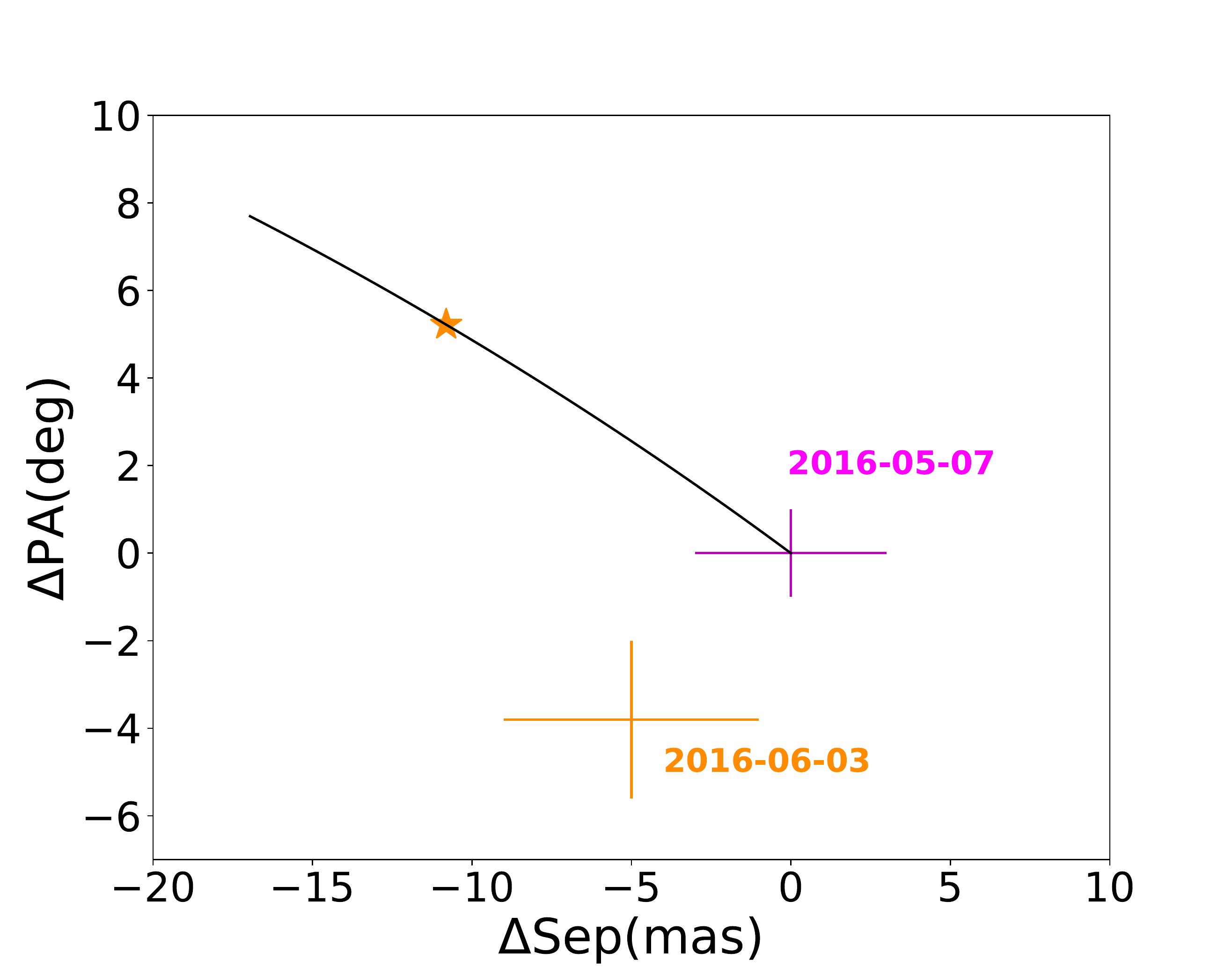}}
  \caption{Astrometry of the candidate companion to the $\lambda$ Muscae binary showing common proper motion. The magenta cross indicates the position of the candidate in the May epoch, taken as ($\Delta$Sep($\arcsec$), PA($\deg$)) = (0,0). The black line shows the trajectory followed by a background object with time. The orange star specifies the location of the candidate in the June epoch if it moved with the background.  The orange cross shows the location at which it is actually found. }
  \label{astro}
\end{figure}

The relative astrometry for the candidate 
is obtained from the combination of our IRDIS-PCA reductions and the $\textit{SpeCal}$ pipeline with cADI and TLOCI, the latter via the injection of a model planet into the data followed by a minimization of the position and flux residuals within a disk of diameter 3 FWHM (Galicher et al. 2018, submitted). For the dataset acquired in May, our IRDIS results are compatible with that obtained with IRDIS-$\textit{SpeCal}$. For the June dataset we did not use the IRDIS-$\textit{SpeCal}$ epoch and thus only used the result given by our own pipeline. The cADI and TLOCI IFS-$\textit{SpeCal}$ astrometry also turned out to be consistent. We adopt the average of the IRDIS and IFS astrometry as final values for each epoch, and their scatter as error bars. These results are shown in Figure \ref{astro}, ruling out background motion with a time span between observations of only one month. This has been possible because the remarkably small separation at which the companion is detected, and the rather high proper motion of the central binary \citep{vanleeuwen2007}. We did not include the ambiguous 2.5\,$\sigma$ January detection, indicated with a question mark in Figure \ref{triple}. Nevertheless, its addition does not alter the fact that the companion is comoving.\\

\begin{figure}
  \resizebox{\hsize}{!}{\includegraphics{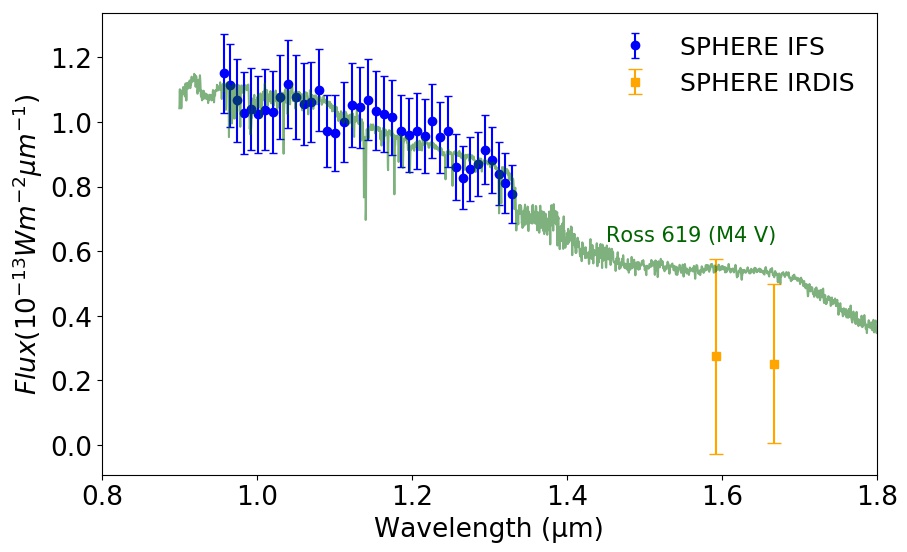}}
  \caption{Flux-calibrated spectrum of the found circumbinary close companion to the binary $\lambda$ Muscae binary system. The template spectrum of the best-fit M4-type Ross 619 star is overplotted for comparison.}
  \label{spectrum}
\end{figure}

To assess the nature of the discovered object, we first obtained its flux-calibrated spectrum between 0.95 and 1.7\,$\mu$m using the contrast of the companion in the IFS and IRDIS bands given by $\textit{SpeCal}$. To this aim, we built the host star SED from several photometric measurements found in the literature. To convert from magnitudes to fluxes, we used the zero points and bandpasses reported in \citet{mann2015}. We then scaled and fit a BT-NextGen \citep{allard2012} synthetic stellar spectrum model ($T\rm_{eff} = 8200\,K, log(g) = 4.0, [Fe/H] = 0$) to the observed flux values. The choice of the model spectrum parameters was based on the atmospheric parameters estimated by \citet{david2015} for the A-type primary component of the $\lambda$ Muscae system. We finally applied the observed companion-to-star contrast ratios, obtained from $\textit{SpeCal}$ for each spectral channel, to convert the primary flux to the flux-calibrated $\lambda$ Muscae C spectrum shown in Figure \ref{spectrum}. The absence of clear absorption bands and the fact that it steadily gets fainter at longer wavelengths, seems to indicate a stellar nature. We note that the location of the object in the speckle-dominated regime imposes significant uncertainties on the error bars; the relative errors are up to 10\,\% and 80\,$\%$ for the IFS and IRDIS bands respectively. \par
\begin{figure}
  \resizebox{\hsize}{!}{\includegraphics{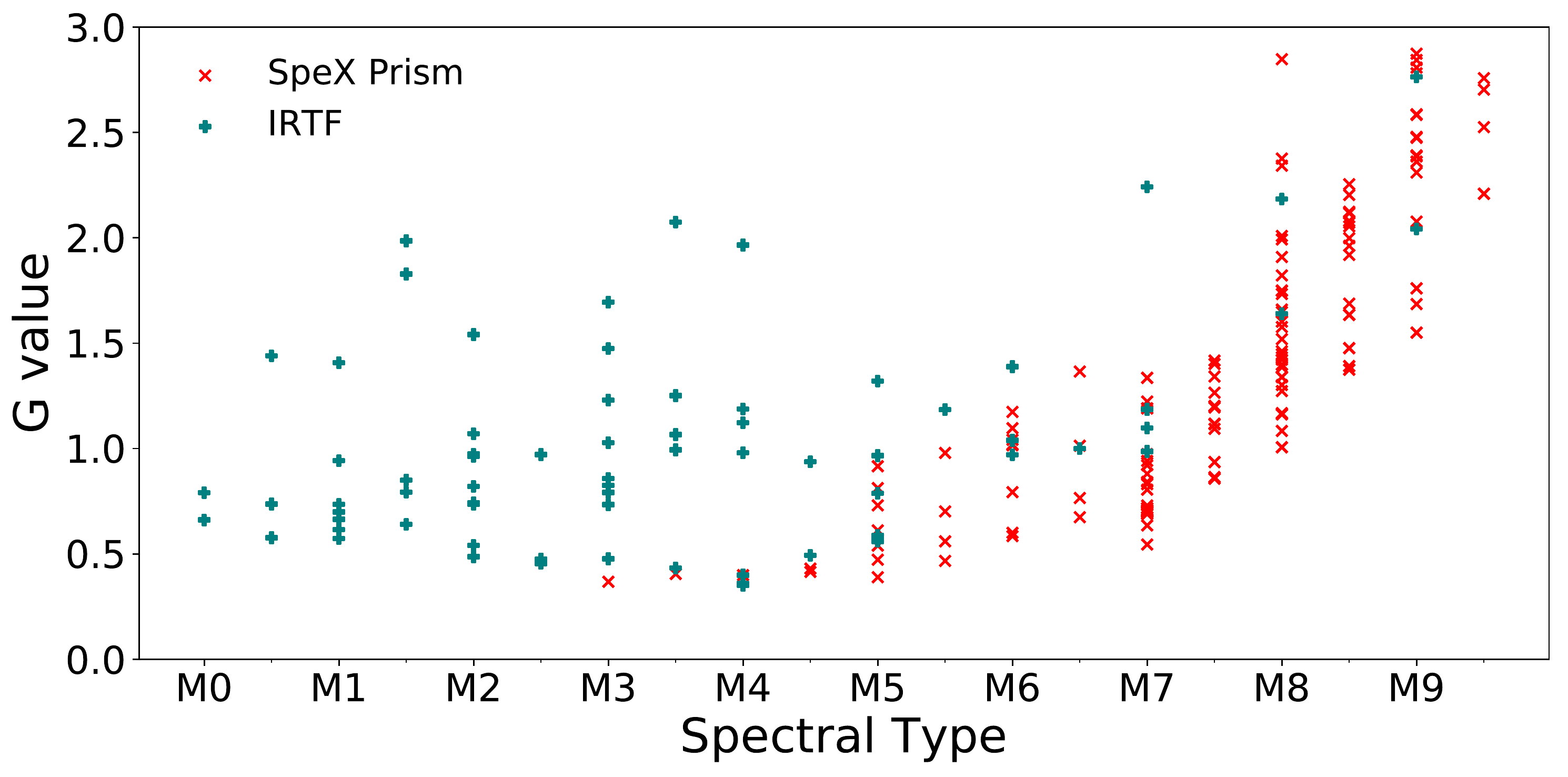}}
  \caption{Goodness of fit statistic results for the comparison between the $\lambda$ Muscae C extracted spectrophotometry and a library of stellar templates obtained with the SpeX spectrograph. Blue and red dots represent data coming from the IRTF \citep{rayner2009} and SpeX \citep{burgasser2014} libraries, respectively. }
  \label{comparison}
\end{figure}

We compared the extracted $\lambda$ Muscae C spectrum with the SpeX and IRTF libraries of reference near-IR stellar spectra obtained with the SpeX spectrograph \citep{rayner2009,burgasser2014}. We adopt the commonly used G goodness-of-fit statistic \citep{cushing2008} to fit the spectrophotometric $\lambda$ Muscae C data to each model. This approach gives a weight to each filter based on its width. The results are shown in Figure \ref{comparison}. The comparison with all the models in the library suggests that  $\lambda$ Muscae C has a spectral type (minimum G value) that lays in the range M2-M6. The best-fit model is the M4-type single star Ross 619 from the IRTF library. We also show its scaled spectrum in Figure \ref{spectrum} for comparison. The two IRDIS bands seem to be slightly offset from the best-fit spectrum. This might also be caused by the location of the companion, which is dominated by speckle noise. As the G-value statistic might indicate that some higher mass stars are able to provide a similar fit to the $\lambda$ Muscae C spectrum, we also derived a mass estimate from the IRDIS H23 flux values. For a $\lambda$ Muscae age of 820\,Myr, the observed contrasts give absolute magnitudes for the companion of M$\rm_{H2}$$\sim$8.7 and M$\rm_{H3}$$\sim$8.6, which yield a mass of $\sim$140\,$M\rm_{Jup}$ and a temperature of $\sim$3000\,K, both in the range of M-type stars, using the evolutionary models from \citet{baraffe2015}.\\

\section{Statistical analysis}
\label{sec:statanaly}
In this section we present the statistical analysis of the results of the SPOTS survey (or SPOTS III), evaluating its impact on the current knowledge of the frequency of substellar companions in circumbinary configuration. 
In order to put more stringent constraints on the matter, we also performed the same analysis for an extended sample of 163 pairs obtained combining our survey with the sample presented in the SPOTS II paper.

\subsection{SPOTS III}

Although our survey did not yield any bona fide substellar companion, we can estimate an upper limit on the fraction of stars having these type of objects from the 5$\sigma$ detection limits shown in Figure \ref{contrast}. We adopt the formalism presented in \citet{laf2007}, which was used for this purpose in several previous works, including SPOTS II. \par
To assess the fraction of binary stars \emph{f} that have at least one companion in the mass and semimajor axis range [m$\rm_{min}$,m$\rm_{max}$] and [a$\rm_{min}$,a$\rm_{max}$], we first define the probability \emph{p$_{j}$} of finding the companion  around a given binary \emph{j}, provided that the companion is actually there. Thus, the probability of detecting and not detecting the companion is simply \emph{fp$_{j}$} and (1 - \emph{fp$_{j}$}), respectively. The likelihood of the set of detections \{d$_{j}$\} given \emph{f} is

\begin{equation}
    \mathcal{L} ( \{d_{j}\}\,\vert\,f) = \displaystyle \prod_{j=1}^{N} (1 - fp_{j})^{(1 - d_{j})} \,(fp_{j})^{d_{j}}
\end{equation}

where \emph{d$_{j}$} = 1 if there is a detection around the star \emph{j}, and 0 otherwise. Now, using Bayes' theorem it is possible to calculate the probability density of \emph{f} given our results as follows:
\begin{equation}\label{eq:posterior}
    p(f\,\vert\,\{d_{j}\}) = \frac{ \mathcal{L} ( \{d_{j}\}\,\vert\,f) p(f)}{\int_{0}^{1}  \mathcal{L} ( \{d_{j}\}\,\vert\,f) p(f) df}
\end{equation}

We simply assume an uninformed prior $p(f)$ = 1. A confidence interval for \emph{f} can be obtained from the posterior for a given confidence level $\alpha$
\begin{equation}
    \alpha = \int_{f\rm_{min}}^{f\rm_{max}} p(f\,\vert\,\{d_{j}\}) df
\end{equation}

\begin{figure}
  \includegraphics[width=11cm]{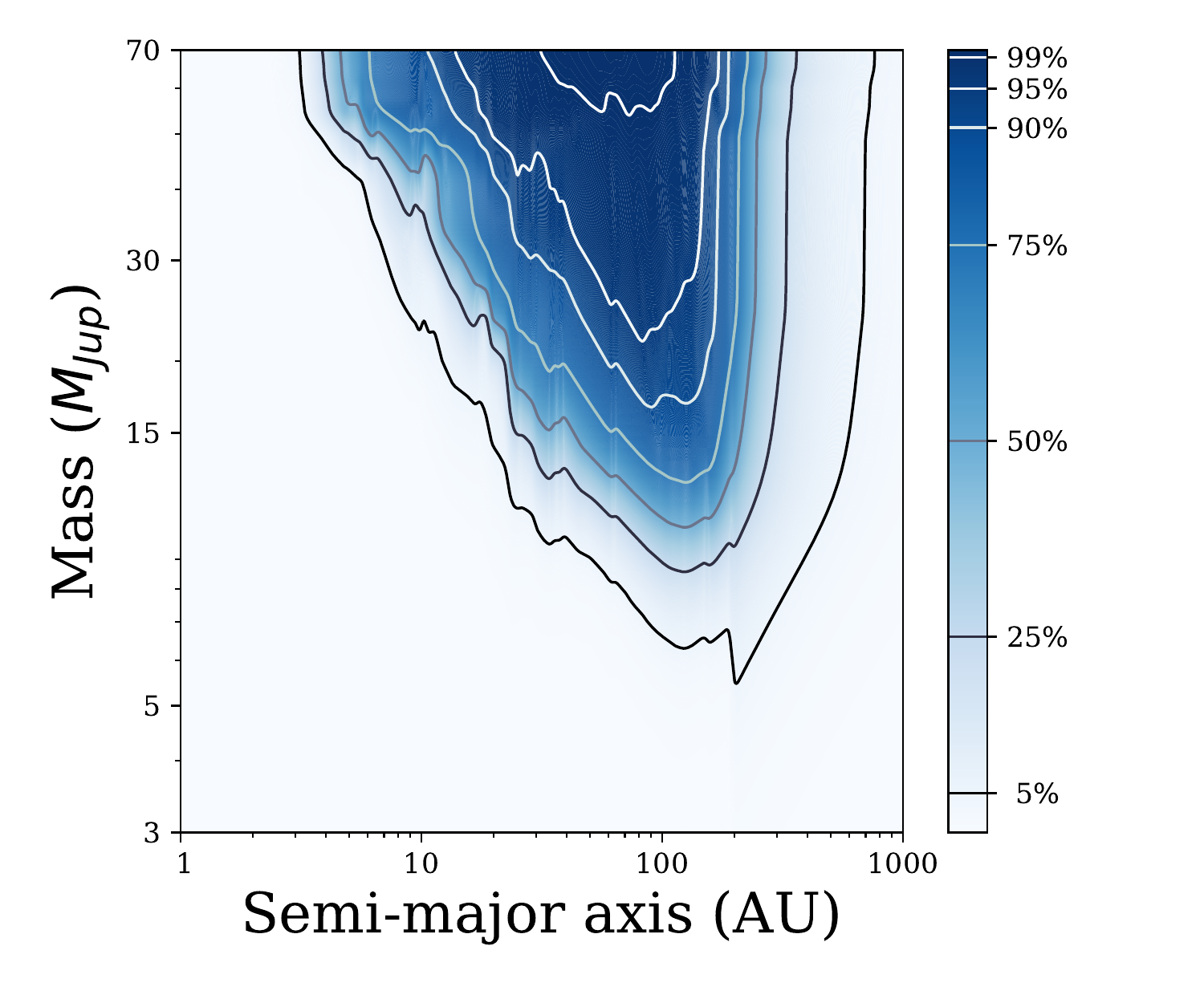}
  \caption{Median detection probability for the SPOTS III sample, dependent on the mass and semimajor axis. The grid used for the code was up to 1000 AU, which we show here. }
  \label{med_det}
\end{figure}

For the non-detection cases, such as our own, f$\rm_{min}$ = 0, and we can easily constrain f$\rm_{max}$ via the simple analytical expression
\begin{equation}\label{fmax}
    f\rm_{max} \approx \frac{-ln(1 - \alpha)}{N\langle\,p_{j}\,\rangle} 
\end{equation}
where $\langle$\,p$_{j}$\,$\rangle$ is the average detection probability in the specified mass and semimajor axis range and $N$ is the total number of binaries. The median detection probability map for our N $=$ 62 binaries is shown in Figure \ref{med_det}. 
We explored semimajor axes up to 1000\,AU, but only the inner 300\,AU are used to compute companion frequencies, which is interior to any of the sources of unknown companionship reported in Table \ref{unknowns}, and thus the results will be valid for the future in any case. The detection probability map has been evaluated via the QMESS code \citep{bonavita2013} with uniform distributions for both semimajor axis and mass; this code uses the contrast curves from Figure \ref{contrast} and the stellar information reported in Table \ref{tab:mastertable}. To convert from contrasts to minimum detectable masses, we made use of the COND and BHAC15 models \citep{allard2001,baraffe2003, baraffe2015}. When more than an epoch was obtained for a given target, the best sensitivity among the curves was considered for each separation. \par
\begin{figure}
  \resizebox{\hsize}{!}{\includegraphics{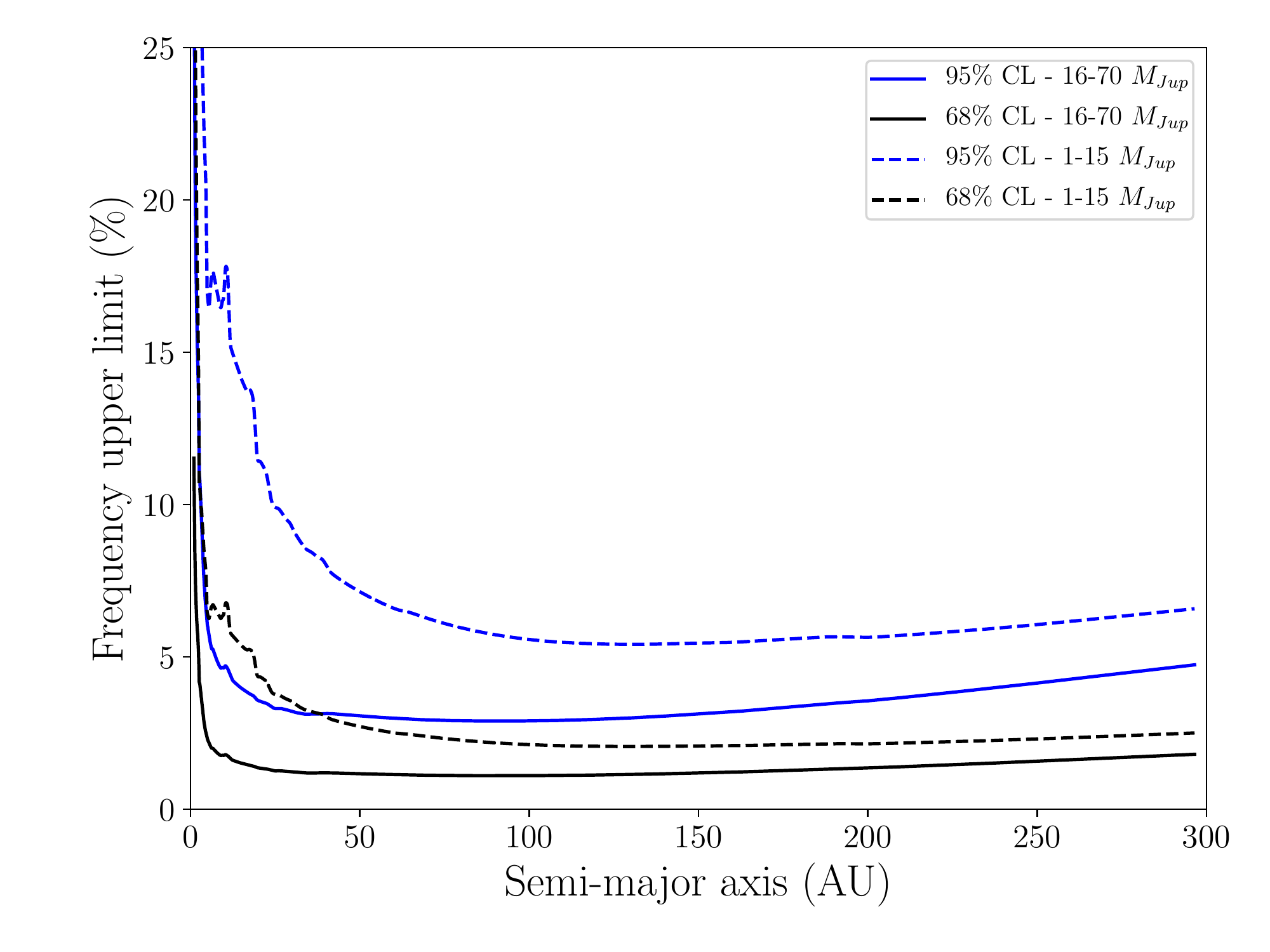}}
  \caption{Maximum frequency of circumbinary companions in our SPOTS III survey as a function of mass and semimajor axis. Constraints on the presence of planetary-mass companions and BDs are shown for a 68\,$\%$ and a 95\,$\%$ confidence level. }
  \label{freq_det}
\end{figure}

Figure \ref{freq_det} shows the resulting upper limit on the companion frequency f$\rm_{max}$, calculated through Eq. \ref{fmax}, that is compatible with the SPOTS III observations. In this way, for a certain range of semimajor axes, the upper limit on the frequency  of companions for a given confidence level is the minimum of the corresponding curve within that interval. For planetary-mass companions of 1--15\,M$\rm_{Jup}$, our data is compatible with a maximum frequency of $\sim$7\,$\%$ at distances beyond  50\,AU at a 95\,$\%$ confidence level. This value goes down to $\sim$2.5\,$\%$ at a 68\,$\%$ confidence level. Broadly, our SPOTS III survey reveals that CBPs are not expected to be more common than $\sim$10\,$\%$ from 30 to 300\,AU of the central binary. Closer distances are difficult to assess given our low sensitivity to planets in those regions. The detection probability of brown dwarfs (BDs; 16--70\,M$\rm_{Jup}$) in our survey is higher than those for planets, and closer distances can be revealed. We find a maximum frequency of circumbinary BDs of only $\sim$2\,$\%$ and $\sim$5\,$\%$ in the range 5--300\,AU for a 68\,$\%$ and a 95\,$\%$ confidence level, respectively.\par
As our survey is sensitive to CBP masses below 15~$M_{Jup}$, there is the possibility that these objects did not form within the protostellar disk, but as the result of direct core fragmentation in a multiple system. Although at these mass ranges the initial mass function is poorly constrained, numerical simulations by \citet{bate2012} show that planetary-mass objects formed via the direct collapse of the molecular cloud should be very rare in general.

\subsection{SPOTS II + SPOTS III}

We combine the SPOTS II archival sample compiled in \cite{bonavita2016} with the SPOTS III targets, forming a total of 163 individual binaries (with
no overlap). Histograms showing the properties of this 
sample can be found in Figure \ref{histogram_IIandIII}. For the  16 overlapping targets, the best observation in terms of detection probability was selected, usually from SPOTS III. As discussed in Section \ref{sec:stell_param}, the determination of the stellar parameters was performed homogeneously in SPOTS II and in the present work, allowing us to merge the samples without specific adjustments.\par
In SPOTS II 5 circumbinary systems with substellar companions were reported. Two of these, HD 106906 b \citep{bailey2014} and 2M0103(AB)b \citep{delorme2013} (see Table 3 in SPOTS II), are in the planetary-mass regime. The properties of 2M0103(AB)b were since revisited by \citet{janson2017} and an alternative mass range of $\sim$15--20\,M$\rm_{Jup}$ was suggested. We however still consider this as an ambiguous case and include it as a  planetary-mass companion.\par
\begin{figure*}[htbp]
\centering
\label{fig:plfreq_cbin}\includegraphics[width=9cm]{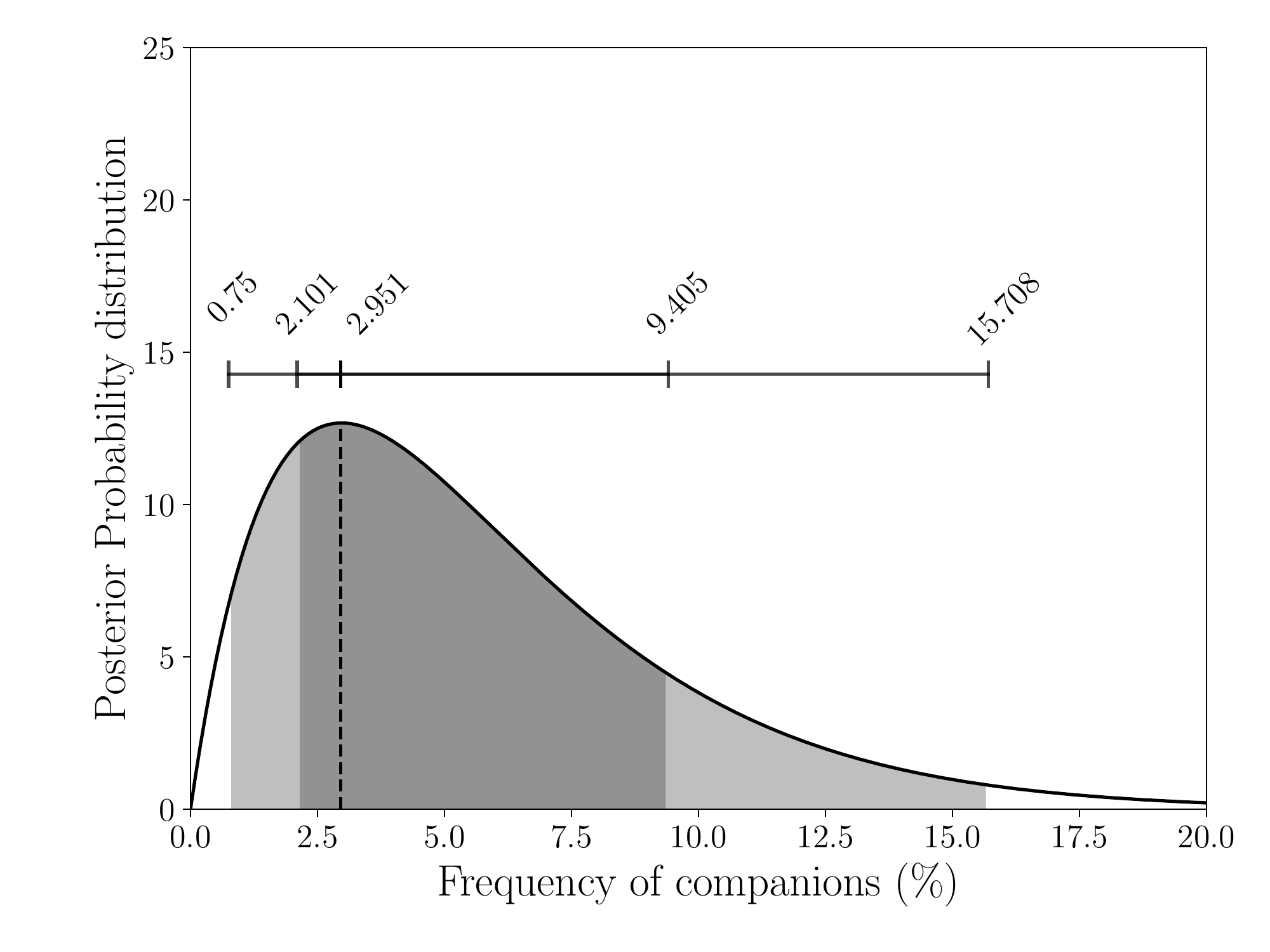}\hspace{0.1em}%
\label{fig:freq_cbin}\includegraphics[width=9cm]{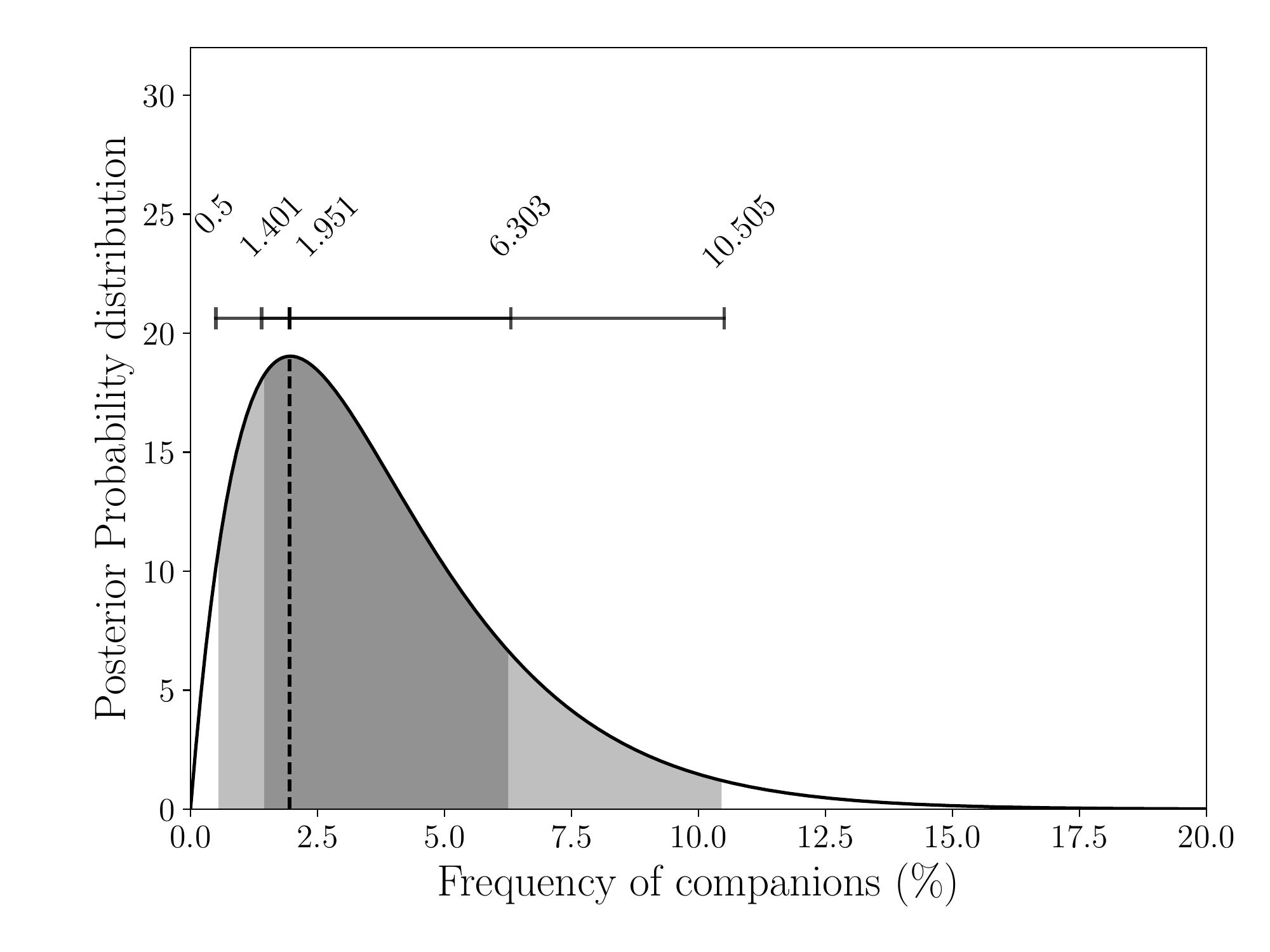}
\label{fig:plfreq_ss}\includegraphics[width=9cm]{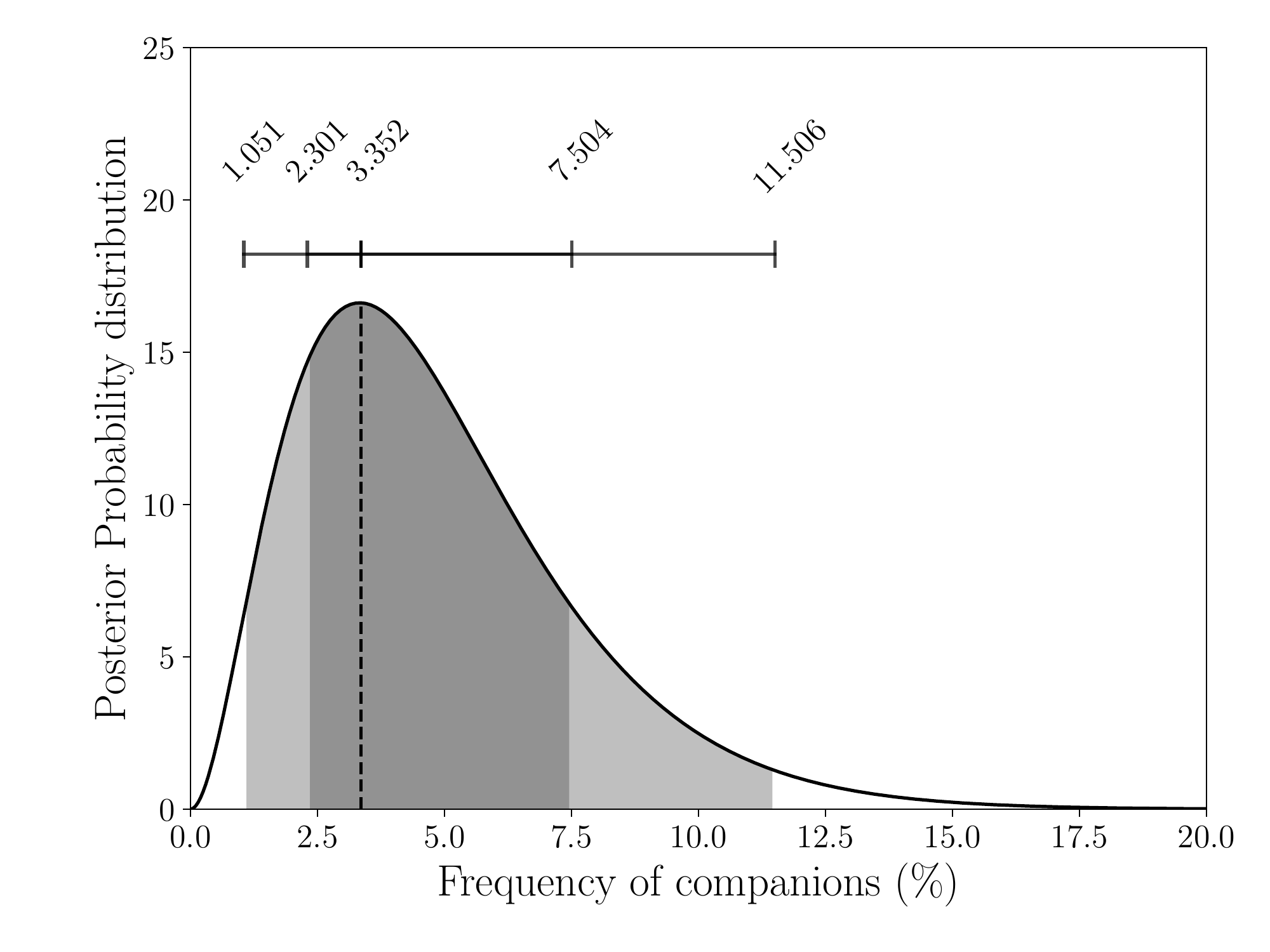}\hspace{0.1em}%
\label{fig:freq_ss}\includegraphics[width=9cm]{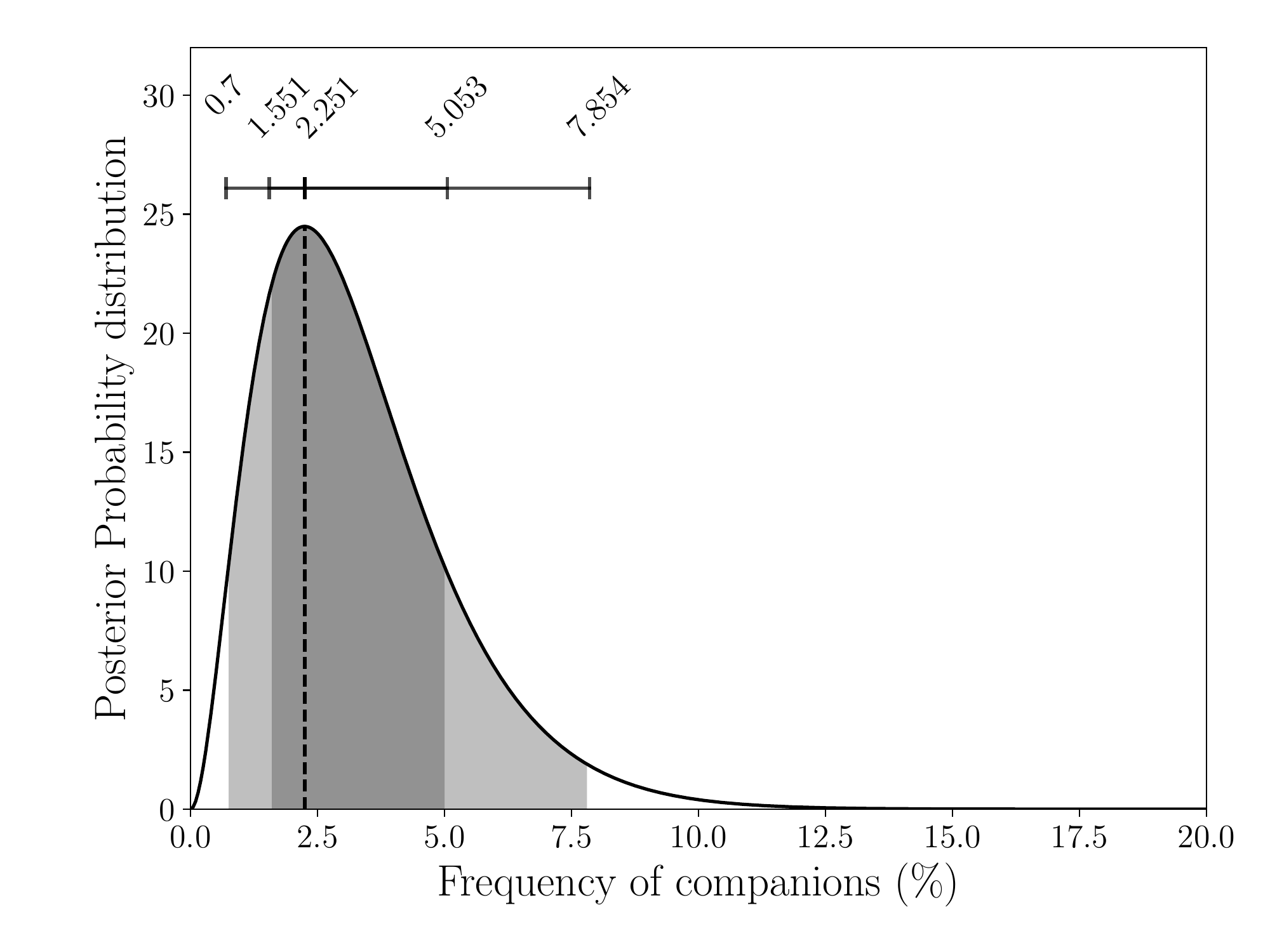}
\caption{Posterior probability distribution of the frequency of planetary mass (2--15~$M_{Jup}$, top panels) and BDs (16-70~$M_{Jup}$, bottom panels). The results of the SPOTS II dataset alone and the combined 163 SPOTS II + III targets are shown on the left and right columns, respectively. The 95\% (light grey) and 68\% (dark grey) confidence levels are shown as shaded areas.}
\label{fig:posteriors}
\end{figure*}

Joining the SPOTS II dataset with our homogeneous sample of 62 binaries, we can use Eq. \ref{eq:posterior} to derive the posterior probability distribution p($f$\,$\vert$\,\{d$_{j}$\}) of the circumbinary companion frequency \textit{f} for the N = 163 binaries within 300\,AU. We left out the two companions found in SPOTS II at separations $\geq$300\,AU, HIP 59960 b and HIP 19176 B. The results are shown in Figure \ref{fig:posteriors} for the planetary (2--15\,M$\rm_{Jup}$) and the BD (16--70\,M$\rm_{Jup}$) cases. The figures extend from 1 to 300\,AU, and show \textit{f} for confidence levels of $\alpha$ = 68\,$\%$ and 95\,$\%$.\par
The combined SPOTS II + III best-fit frequency for CBPs is 1.95\,$\%$, with a 10.50\,$\%$ upper limit at a 95\,$\%$ confidence level. This is in line with a frequency of 2.95\,$\%$ found for the SPOTS II dataset alone (see left panels in Figure \ref{fig:posteriors}) and is compatible with the upper limits derived in Figure \ref{freq_det} for SPOTS III. In the range of BDs, we find a similar SPOTS II + III best-fit value with a 2.25\,$\%$ frequency, but a lower upper limit of 7.85\,$\%$. A slightly higher value of 3.35\,$\%$ is found for the SPOTS II targets, which is however consistent within the 68\,$\%$ confidence level.\par

\subsection{Comparison to other CBPs occurrence rates and massive planets around single stars}

The most common source of CBPs population studies is the Kepler survey, due to its large sample size and good sensitivity to such companions.
\citet{welsh2012} found a 2.8\,\% frequency of short-period planets based on an analysis of 750 eclipsing binary systems. \citet{Martin2014} simulated circumbinary distributions of planets around non-eclipsing binaries and compared these to Kepler detections, estimating a minimum occurrence rate of about 9\,$\%$, which agrees well with the result of \citet{armstrong2014} (see \citet{martin2018} for a review on CBPs population). These results are complementary to our SPOTS survey, as our targets are younger and the sensitivity space is different than Kepler's observations. \par
A trend that appears to have emerged from these indirect observations is that
CBPs are piled up at the edge of the dynamical stability limit \textit{a$\rm_{c}$}, which is attributed by some authors to an inward migration, halting when the orbit becomes unstable \citep[e.g.,][]{kley2014}. However, numerical simulations show that more than half of the CBPs found by Kepler could contain another equal-mass planet closer to the circumbinary stability radius, which combined with the bias that the transit method has toward small orbits, seems to dismiss this pile-up near the stability limit \citep{quarles2018}. In our survey, the majority of the \textit{a$\rm_{c}$} are located at $\leq$\,20\,AU (see Figure \ref{histogram}), which from Figure \ref{med_det} corresponds to median detectability probability values of less than $\sim$10\,$\%$ for our planetary-mass range. If these planets indeed accumulate near \textit{a$\rm_{c}$}, even though our survey is the most extensive direct imaging survey to date resolving close distances to binary stars, it might not be sensitive enough to recover the population of planets located at those distances. In this way, although the low-mass companion presented in Section 4.4 around $\lambda$ Muscae is not a planetary-mass object, it is precisely located at a distance where we expect \textit{a$\rm_{c}$} to reside.\par
As done in SPOTS II, we can compare our very low frequency results of CBPs to the frequency of post-common envelope binaries having signs of period variations. \citet{zorotovic2013} found that $\sim$90\,$\%$ of these systems show apparent period variations, which might be caused by massive planets at distances that would be detectable by our survey. If these variations happen to be caused by the presence of actual planets, this favors a second-generation origin of planets around post-common envelope binaries.\par
Finally, the SPOTS statistics around binary stars can be compared to previous direct imaging surveys focusing on young and nearby single stars, which are sensitive to the mass ranges and distances that we are probing in this work. For more than a decade, direct imaging surveys have been targeting tens or hundreds of these stars, gaining in contrast as better AO systems were developed and novel post-processing techniques were employed \citep[e.g.,][]{bowler2016}. All these substellar detections and sensitivity curves have been merged to find a planetary-mass frequency of about 1\,$\%$, averaged across spectral types in the range 5--13\,M$\rm_{Jup}$ at 5--500\,AU \citep{bowler2018}. This number seems to increase slightly for BDs (13–-75\,M$\rm_{Jup}$) up to about 1–-4\,$\%$. Better constraints and trends with stellar mass or the presence of debris disks are expected in the near future, as the surveys conducted by the second-generation AO instruments, such as SPHERE \citep{chauvin2017} and GPI \citep{macintosh2015} targeting about 600 stars each, are in their last phase of observations.

\section{Conclusions}
We have presented the first direct imaging survey dedicated to finding  CBPs. We analyzed a total of 62 close binaries with VLT/NaCo and VLT/SPHERE, whose main results can be summarized as
\begin{itemize}
    \item No substellar companion has been found around any of the 62 binaries inside 300\,AU, although there are a few interesting candidates lacking follow-up observations further out. We also presented the resolved circumbinary disk around AK Sco \citep{janson2016} and the discovery of a low-mass star orbiting the $\lambda$ Muscae binary at a separation of only $\sim$0.15\,$\arcsec$.
    \item This non-detection gives an upper limit on the frequency of CBPs (1--15\,M$\rm_{Jup}$) and BDs (16--70\,M$\rm_{Jup}$) of $\lessapprox$10\,$\%$ and $\lessapprox$6\,$\%$, respectively, in the range $\sim$30--300\,AU at a 95\,$\%$ confidence level.
    \item Including the archival SPOTS II data, we analyze a total of 163 binary systems; we find a best-fit CBP (2--15\,M$\rm_{Jup}$) frequency of 1.95\,$\%$ with an upper limit of 10.50\,$\%$, and a
    2.25\,$\%$ frequency for the BD (16--70\,M$\rm_{Jup}$) case with a  7.85\,$\%$ upper limit at a 95\,$\%$ confidence level.
    \item These values are very similar to the occurrence rate of giant planets and BDs in wide orbits around single stars in the last surveys, converging to about $\sim$1\,$\%$ and $\sim$1--4\,$\%$, respectively (see \citet{bowler2018}).

\end{itemize}

We have proven that, with the right selection criteria, binary stars do not imply a detriment to the high-contrast imaging technique. Bigger samples will be needed to better constrain the occurrence rates of CBPs, and further observations will find out the unknown companionship of some of the wider SPOTS candidates, such as the three objects around HIP 77911. The next generation of extremely large telescopes will be ideally suited for reaching the critical radius of stability and probe inner regions where planets might reside and even accumulate. In addition, SHINE \citep[e.g.,][]{chauvin2017} and the Gemini Planet Imager Extra Solar Survey \citep[GPIES;][]{macintosh2015} will soon provide a significantly improved comparison sample for single stars. The GAIA mission will help to reveal the presence of massive planets at close separations, although the impact of binarity is uncertain, and will provide new targets for a circumbinary search. the Transiting Exoplanet Survey Satellite (TESS) is also expected to provide hundred of CBPs similar to those found by Kepler \citep{quarles2018}.

 \begin{acknowledgements} 
R. Asensio-Torres and M. Janson gratefully acknowledge funding from the Knut and Alice Wallenberg foundation. S.Desidera acknowledges support from the "Progetti Premiali" funding scheme of the Italian Ministry of Education, University, and Research. This work has been supported by the project PRIN-INAF 2016 The Cradle of Life- GENESIS- SKA (General Conditions in Early Planetary Systems for the rise of life with SKA).
 This work has made use of the SPHERE Data center, jointly operated by OSUG/IPAG (Grenoble), PYTHEAS/LAM/CeSAM (Marseille), OCA/Lagrange (Nice) and Observatoire de Paris/LESIA (Paris) and is supported by a grant from Labex OSUG@2020 (Investissements d’avenir – ANR10 LABX56). This research has made use of the NASA Exoplanet Archive, which is operated by the California Institute of
Technology, under contract with the National Aeronautics and Space Administration under the Exoplanet Exploration Program. 
The distortion correction of the HiCIAO images is based on observations made with the NASA/ESA Hubble Space Telescope, and utilized the data obtained from the Hubble Legacy Archive, which is a collaboration between the Space Telescope Science Institute (STScI/NASA), the Space Telescope European Coordinating Facility (ST-ECF/ESA), and the Canadian Astronomy Data center (CADC/NRC/CSA). Our data analysis benefited from PyRAF and PyFITS that are the products of the STScI, which is operated by AURA for NASA. Data analyses were in part carried out on common use data analysis computer system at the Astronomy Data Center, ADC, of the National Astronomical Observatory of Japan.

 \end{acknowledgements}

\clearpage

\onecolumn
\setcounter{table}{3}

\small{
\tiny{
\begin{landscape}
\begin{longtable}{ l l l c c c c c c c c c c c}
\caption{\label{tab:mastertable} SPOTS targets } \\
\hline\hline 
  Target ID            &  RA(2000)   &  Dec(2000)   & Dist  &  Age   &    H    &   SpT  & M$\rm_A$     &  M$\rm_B$  &  $\rho$     &  P     & ecc  & $a_{crit}$ & Notes  \\
             &             &              &  (pc) & (Myrs) &  (mag)  &        & ($M_{\odot}$)  & ($M_{\odot}$) &  (mas) & (days) &      &  (AU)   &      \\   
\hline
\endfirsthead
\caption*{Table 4, continued.}\\
\hline\hline
  Target ID            &  RA(2000)   &  Dec(2000)   & Dist  &  Age   &    H    &   SpT  & M$\rm_{Prim}$     &  M$\rm_{Comp}$  &  $\rho$     &  P     & ecc  & $a_{crit}$   & Notes  \\
            &             &              &  (pc) & (Myrs) &  (mag)  &        & ($M_{\odot}$)  &  ($M_{\odot}$) &  (mas) & (days) &      &  (AU)   &        \\   
\hline
\endhead

\hline
\endfoot

\hline 
\caption*{}
\endlastfoot

HIP 7601\phantom{\Large!}= HD 10800  & 01:37:55.56 & -82:58:29.98 & 26.7 & 600 & 4.53 & G1V & 1.104+0.929 & 1.033 & 78 & 638.24 & 0.19 &  0.62 & Triple  \\
HIP 9892 = HD 13183   & 02:07 18.10 &  -53:11:56.00   &   49.35     &   45       &   6.98     &   G7V        &    1.0   &   --     &    --        &    --     &  --     &     $<$ 19 &      \\ 
HIP 12225 = $\eta$ Hor  & 02:37:24.37 & -52:32:35.08 &  41.9 & 797 & 4.6  & A6V + F0V & 1.75 &  1.65 &  68.7 & 1098.65 & 0.16 &  8.65  &\\
HIP 12545  = BD+05 378      & 02:41:25.90 &  +05:59:18.40   &  44.4       &   24       & 7.20   &   K6V   & 0.76     &    --    &   --       &   --    &   --  &   $< $16   &  \\ 
HIP 12716 = UX~For                   & 02 43 25.57 &  -37:55 42.53   &  41.1      &  600       & 6.10   &  G6V     & 0.88+0.64 &  0.58       &  305.30     &   --    &   --  &   57.5  & Triple      \\ 
HIP 14007 = HD 18809  & 03:00:19.71 & -37:27:16.17 & 48.9 & 100 & 6.9 & G3V + K5V & 1.05 & -- & -- & -- & -- &  $<$ 19 & \\
HIP 14568 = AE For  & 03:08:06.66 & -24:45:34.74 & 32.2 & 4000 & 6.85 & K7V & 0.6314 & 0.6197 & 0.6 &  0.92 & 0.00 & 0.05 &  \\
HIP 14807 = BD+21 418B  & 03:11:12.33 & 22:25:22.73 & 56.2 &  149 & 8.1 & K6 & 0.76  & 0.52 & 28.74 & -- & -- &  5.63 &\\
HIP 15197 = $\zeta$ Eri  & 03:15:50.02 & -08:49:11.02 & 35.3 & 800 & 4.22 & A5 & 1.77 & --& -- & 17.93 & 0.14 &   0.52  &\\
HIP 16853 = HD 22705     & 03:36:53.40 &  -49:57:28.90   &  45.3       &   45       & 6.26   &  G2V      &  1.0      &  0.40       &   --       &  201.0  & 0.00  &   1.77  &    \\ 
HIP 19591  = V1136~Tau = HD 284163                  & 04 11 56.22 &  +23:38:10.76   &  40.3      &  625       &   $\sim$7.1     &  K0       &0.78+0.49  &  0.52       &   310     &  14928.5    & 0.853 &  51.1 & Triple  \\  
TYC 5907-1244-1            & 04:52:49.50 &  -19:55 02.00   &  61.7       &   45       & 7.50   &  K1V     &  0.87     &     --        &   --       &   --     &   --  &   $<$ 23 &  \\ 
HIP 25486 = AF~Lep         & 05:27:04.76 &  -11:54:03.47   &  26.9    &   24       & 2.93   &  F7V      &  1.06    & 0.76    &   --       &   --     &   --  &   $<10$ &     \\ 
HIP 25709 = HD 36329         & 05:29:24.10 &  -34:30:56.00   &  72.5      &   42       &    7.06    &  G3V      &  1.07    &      --       &    --        &     --     &    --   &    $<$ 27  &   \\ 
HIP 27134 = XZ~Pic            & 05:45:13.40 &  -59:55:26.00   &  47.9       &  300       &    7.1    &  K0V   &  0.95    &   --        &   --       &   --     &   --  &   $<$ 18    & \\ 
HIP 31681 = Alhena = $\gamma$~Gem & 06:37:42.71 &  +16:23:57.41   &  33.5       &  300       &  1.84  &  A1.5IV  & 3.0      &     1.0        &  273    &  4609.95 & 0.89  &     40   &    \\ 
TYC 8104 0991 1            & 06:39:05.70 &  -45:12:58.00   &  98.3       &  100       & 8.25   &  K3V     & 0.83 + ?      &    0.78         &    133.0      &     --     &   --       &    $<$ 38   &  Triple   \\ 
HIP 32104 = 26~Gem                & 06:42:24.33 &  +17:38:43.11   &  47.5     &   42       & 5.07   &  A2V      & 2.70      & 0.51       &   --       &    522   &  0.10 &  4.76    &   \\  
EM Cha = RECX7                 & 08:43:07.24 &  -79:04:52.50   &  98.7   &    11       & 7.75   & K7V      & 1.0       & 0.4        &   --       &    2.6   &  --   &  0.15   &   \\ 
TYC 8569 3597 1            & 08:51:56.40 &  -53:55:57.00   & 129.0      &   45       &     8.90   &  G9V      &  1.11         &    --    &   --       &   24.06  &  --   &  0.73   &   \\ 
TYC 9399 2452 1            & 09:19 24.00 &  -77:38:45.00   &  68.2     &  800       & 7.56   &  K0IV    & 0.97          &   --          &  --   &    --      &  --      & $<$21 &  wide comp.   \\ 
HIP 46509 = $\tau$ Hya   & 09:29:08.89 & -02:46:08.26 &  17.8  & 2000 & 3.54 & F6V & 1.36  & 0.35  & 341.79 & 2815 & 0.427 & 17.1 &\\
HIP 46637 = GS~Leo                   & 09:30:35.83 &   10:36:06.25   &  35.2        &  300       & 6.69   &   G5      & 0.89      &   --         &   --       &    3.86    &  --   &  & wide comp.   \\ 
HIP 47760 = HD 84323                 & 09:44:14.10 &  -11:52:21.00   &  76.5      &   30       &   7.76     & G3V       &  1.08            &  --        &  --  &   --      &  --      &   $<$ 29 &  \\ 
TYC 6604 0118 1    & 09:59:08.40 &  -22:39:35.00   &   60.2      &  100           & 7.49   & K2V      &   0.83   &    0.74  &  --  &   1.84    & 0.00         &  0.08   &       \\ 
HIP 49669 = $\alpha$ Leo = Regulus  & 10:08:22.31 & 11:58:01.95 & 23.8 & 600 & 1.66 & B7V & 3.4 & 0.3 & -- & 40.11 & 0.00 & 0.67 &  wide comp.  \\ 
HIP 49809 = HD 88215  & 10:10:05.88 & -12:48:57.32 & 27.8 & 800 & 4.46 & F3V & 1.41 & 0.2 & -- & 28.1& 0.07&0.49 &  \\ 
HIP 56960 = HD 101472  & 11:40:36.58 & -08:24:20.36 & 43.5 & 300 & 6.15 & F8.5V & 1.2 & $\geq$0.15 &  -- & 4.4 & -- & 0.18 &   \\ 
HIP 57363 = $\lambda$ Muscae  & 11:45:36.41 & -66:43:43.54 & 39.2 & 820 & 3.28 & A7 & 2.06 + 0.39  & 0.16 & 153.0 &  -- & -- & 5.6  & \\
TYC 9412-1370-1   & 12:20:34.36 & -75:39:28.72 & 70.4 & 50 & 8.1 &K3V &0.69 & -- & -- & 613.9 &0.22 & 4.7  &\\ 
HIP  63742  = PX Vir  &  13:03:49.65 &  -05:09:42.50   &  20.5       &   149    & 5.67  & G5V     & 0.84    & 0.51   &  34     & 216.9   & 0.30     &    2.43  &\\  
HIP 69781 = V* V636 Cen & 14:16:57.91 & -49:56:42.36 & 71.5 & 250  & 7.12 & G0V + G7V & 1.05 & 0.85 & -- & 4.28  & 0.13 & 0.18 & \\
HIP 74049  = HS~Lup         & 15:07 57.75 &  -45:34:45.84   &  41.9       &  250       &    6.1    & G5IV+G5IV &   1.10           &      1.08      &  --  &  17.83    &  0.20    &   0.45   &      \\
1RXS J153557.0-232417  & 15:37:40.05 & -23:06:40.0 & 145 & 11& 9.6 & K3 & 0.99&0.10 &  54.68 & -- & -- & 29.7 &   \\ 
HIP 76629 = HD 133822     & 15:38:57.50 &  -57:42:27.00   &  40.1       &   24       &   5.9     & K0V + M       &    1.10          & 0.29    &  0.08    & 1642.5 & 0.53    & 12.0 & wide comp.  \\ 
HIP 77911 = HD 142315  & 15:54:41.59 & -22:45:58.50 & 161.8  & 11 & 6.7 & B9V &2.9 & $\geq$0.24 & -- &1.26 &0.61 & 0.14 & wide comp.   \\
HIP 78416 = HD 143215     & 16:00 31.32 &  -36:05 16.62   &  103.5       &   16       &     $\sim$7.4   & G1V       &   1.26    &      --      &   -- &   --      &  --      &  $<$ 40 & wide comp.   \\
RX J1601.9-2008 = BD-19 4288  & 16:01:58.22 & -20:08:11.99 &154 & 11 &7.8 & G5 &1.46 & 0.82 &36.18 & 2982.05 & 0.35 & 19.34  & \\
ScoPMS027 = V1156 Sco  & 16:04:47.74 & -19:30:22.92 & 140.27 & 11 & 8.27 & K2IV & 1.12 & 0.74 & 43.18 & -- & --  & 23.7   &\\ 
TYC 6209-735-1  & 16:08:14.74 & -19:08:32.77 & 142.13 & 16 & 8.6 & K2 & 1.14 & 0.45 & 28 & 2015 &0.21 &  11.3  &\\
ScoPMS044 = V1000 Sco = Wa Oph 1  & 16:11:08.90 & -19:04:46.86 & 145 &  11 & 7.98 & K2IV & 0.95  & -- & 4.33 & 144.7 & 0.26 & 2.04 &  \\
ScoPMS048 = V1001 Sco  & 16:11:59.27 & -19:06:53.36 & 145 & 11 & 8.3  & K2 & 1.35 & -- & -- & 10.4 & 0.18 &  0.33 & wide comp. \\
TYC 6213-306-1  & 16:13:18.58 & -22:12:48.99 & 170.6 &  11 & 7.59  & G9 & 1.7 & 1.65  & --  & 166.9 & 0.226 & 2.7 &  \\ 
HD 147808  & 16:24:51.38 & -22:39:32.50 & 140.8 & 11 & 7.28 & G6 &1.54  &1.42 &  58.19& 4851.4 & 0.262 & 25.6  & \\
HIP 80686 = $\zeta$ TrA  & 16:28:28.14 & -70:05:03.84 & 12.2 & 500 & 3.65& F9V + M4V & 1.12& 0.2 & -- & 12.97 & 0.01442 &  0.26  &\\
ROX 33 =  EM* SR 20 & 16:28:32.66 & -24:22:44.90 &120 & 2 & 7.48 &G7 & 3.53 & 0.9 & 48.55 & -- & -- &   23.01 & \\
ROXs 43A   & 16:31:20.12 & -24:30:05.0 & 145.5 & 2 &  7.17  &G5 + K3 & 1.14 + $\geq$0.3 & 0.2  & 0.29  & -- & -- &  193.2 &  Triple + wide comp.  \\
HIP 82747 = AK Sco  & 16:54:44.84 & -36:53:18.56 & 143.5 &  16 & 7.06 & F5+F5 &1.41 &1.39 & 1.11 & 13.6 & 0.47 &   0.56  &\\
HIP 84586 = HD 155555  & 17:17:25.50 & -66:57:03.73 & 30.5  & 24 & 4.91 & K1 & 1.059 & 0.986 & -- & 1.68 & 0.00&0.07  & wide comp. \\ 
HIP 88481 = HD 165045  & 18:04:01.87 & 01:49:56.64 & 32.6 & 600 & 6.23 & G5 & 0.92 & 0.74 & 0.047 & 587.65 & 0.501 & 5.99  & \\
HIP 94050 = HD 177996  & 19:08:50.48 & -42:25:41.18 &  35.8 & 400 & 5.97&K1.5V &0.94 &$\sim$0.7 &-- &-- &-- &$<$ 12   &\\ 
HIP 98704 = HD 188480  & 20:02:50.12 & -75:20:57.50 & 88.5 & 250 & 7.09 & F8 & 1.35 & -- & -- & -- & -- & $<$ 35 & \\
HIP 100751 = $\alpha$ Pav  & 20:25:38.85 & -56:44:06.32 & 54.82 & 45 &  2.46 & B7 &5.82 & 0.26 & -- & 11.7 & 0.0 & 0.32  &  \\ 
HIP 101422 = HD 195289  & 20:33:13.27 & -58:06:44.44 & 106.4  & 100 & 6.41  & G0V & 1.55 & --& -- & -- & -- &  $<$ 1 &\\
HIP 104043 = $\alpha$ Oct  & 21:04:43.06 & -77:01:25.56 & 44.4 & 1100 &3.6 & A7III + G2III & 1.9& -- & -- & 9.073 &0.39 &  0.43  &\\
HIP 105404 = BS~Ind               & 21:20:59.80 &  -52:28:40.10   &  52.7     &   30       & 6.70   & K0V       &  0.8 & $\geq$(0.45 + 0.45)    &   -- &  1204.5     & 0.6     &  11.34  &  Triple  \\ 
HIP 105860 = IK Peg                & 21:26:26.66 &  +19:22:32.30   &  49.72     &  100       &   5.48     & A8    &  1.45        &  0.99      &  --  &  21.72    & 0.00     &     0.45 &      \\
HIP 107556 = $\delta$ Cap          & 21:47:02.44 &  -16:07:38.23   &  11.87       &  540       & 2.01   & A5        &  1.50        & 0.56       &  --  &  1.0     &  0.01     &  0.06    &  \\
HIP 109110 = NT Aqr  & 22:06:05.32 & -05:21:29.06 & 34.05 & 1000 &6.0 &G2V & 1.15& -- & --  &4781.5 & -- &  24.4 & \\
HIP 109901  = CS~Gru                & 22:15:35.20 &  -39:00 51.00   &  53.0    &  100       & 7.12   & K0V       &  0.89        & 0.45       &  --  &   --     &  --       &  $<$ 20 &     \\ 
TYC 6386 0896 1            & 22:44:38.90 &  -15:56:29.00   &  57.9        &  125       & 7.47   & K1V     &  0.88            &     --       &  --  &   --     &  --       &  $<1$  &   \\ 
HIP 113860 = pi. PsA  & 23:03:29.81 & -34:44:57.88 & 29.42 & 175 & 4.51 & F1V + K2 &1.51 & -- &-- &178.31  &0.53 & 3.14 & \\

\end{longtable} 	

\end{landscape} 
}}

\twocolumn
 
\bibliography{ref}

\clearpage

\begin{appendix}
\section{Appendix: Notes on individual targets}
\label{appendix}

\noindent
\textbf{HIP 7601 = HD 10800:} Triple-lined coplanar SB3 system formed by a close SB2 \citep{Wichmann2003}, with masses of 1.104\,\(M_\odot\) and 0.929\, for the primary and secondary, \(M_\odot\),
respectively, and an outer visual and spectral component. A fit to HARPS radial velocity data by \citet{Tokovinin2016} gives an inner orbit with a period of 19.37 days and $e = 0.1$ . The outer AB component was resolved with speckle interferometry by the 4.1\,m SOAR telescope \citep{Tokovinin2015}, which provided the orbital solution. Combining spectral data and visual interferometry, this leads to an outer star of mass 1.033\,\(M_\odot\) orbiting the inner pair every 1.75 years in an orbit with eccentricity $e = 0.19$ and semimajor axis 0.078\,\arcsec. The dynamical mass calculation of the inner pair is then coupled to the spectroscopy to estimate their orbital solution. The system is  an X-ray source, whose chromospheric activity gives an age on the order of 1200\,Myr \citep{maldonado2012}, although the result may be biased from unrecognized multiplicity. Lithium equivalent width (EW) measurements are in concordance to similar  stars in the Hyades, giving an age of 600\,Myr \citep{Tokovinin2016}. Moreover, the spatial velocity of A,B corresponds to the kinematics of the young disk \citep{Tokovinin2016}. Even though the system might be tidally locked, lithium measurements and young disk kinematics clearly point to a younger age. We adopt an age of 600\,Myr. \\
Our SPHERE observation covers only 8 degrees of sky rotation, but detects a candidate low-mass stellar companion at about 2.7\,$\arcsec$ ($\sim$75\,AU). Unfortunately no follow-up epoch was obtained for this target. We also resolve the outer component, previously done in \citet{Tokovinin2015}, at a separation of only 88\,mas (see Table \ref{binaries}).\\

\noindent
\textbf{HIP 9892 = HD 13183:} Long-period SB member of Tucana. See also SPOTS I. Only a likely background galaxy is detected in our first NaCo observation, not redetected in the follow-up.\\ 

\noindent
\textbf{HIP 12225 = $\eta$ Hor:} Astrometric and visual binary, resolved for the first time with speckle interferometry at SOAR by \citet{Hartkopf2012} and more recently by \citet{marion2014} with infrared interferometry at the VLTI. The resolved positions might match the near-circular 3 yr (or photocenter displacement of $\sim$21\,mas) orbit solution, which we adopt here, proposed by \citet{goldin2007} based on Hipparcos stochastic astrometry. \citet{david2015} determined an age of 797\,Myr, which we adopt as it is compatible with kinematics. For this age we derive a primary mass of 1.75\,\(M_\odot\), and used our observed magnitude difference in $H$ band to estimate a mass of 1.65\,\(M_\odot\) for the secondary. \\ 
We only count with one epoch for this target. It suffers from low-wind effect and an average seeing of $\sim$1.6\,$\arcsec$. We detect at a 2$\sigma$ confidence level a candidate companion at $\sim$5.3\,\arcsec, which we would normally not take into account given the low confidence level. However, this is likely the same object as that reported by \citet{ehrenreich2010} in $K$ band with NaCo, and is classified as   of ambiguous nature. Our observations are consistent with the background trajectory.\\
We also resolve the inner binary in the unsaturated frames and the secondary is located at a separation of 73.3\,mas (see Table \ref{binaries}). As mentioned earlier, this binary was previously resolved by two other
works. In January 2011, \citet{Hartkopf2012} measured a separation of 70\,mas and 60\,$\deg$, while \citet{marion2014} redetected it at 78.7\,mas and 40.9\,$\deg$ in December 2012. Our result qualitatively matches well a $\sim$3\,yr orbit.\\

\noindent
\textbf{HIP 12545  = BD+05 378:} Member of the $\beta$ Pic YMG. See SPOTS II. \\

\noindent
\textbf{HIP 12716 = UX~For:} Triple system formed by an SB2 and a resolved outer component in an estimated 40\,yr period orbit.  See SPOTS I for a detailed description of the system. Our NaCo observation resolves the tertiary component, whose astrometry and photometry is reported in Table \ref{binaries}.\\ 

\noindent
\textbf{HIP 14007 = HD 18809:} This SB1 is probably a member of the Octans-Near association (30--100\,Myr). The lithium EW matches the expected value for a G-type star of 100\,Myr, and the activity level from \citet{cutispoto2012} is consistent with Octans-Near age. Based on these observables and the probable association, we adopt an age of 100\,Myr. No orbital solution is available.\\

\noindent
\textbf{HIP 14568 = AE For:} Eclipsing binary of Algol type and SB2; see \citet{Rozy2013} for the complete orbital solution. AE For is tidally locked, which inhibits the use of parameters linked to rotation or activity.
The age of the system based on lithium measurements is controversial, as \citet{Rozy2013} alerted that they could not recover the Li 6708\,$\AA$ EW line of 80\,m$\AA$ that \citet{torres2006} used to estimate an age similar to the Pleiades (125\,Myr). We then derived
U, V, W space velocities, which are  clearly far from the kinematic locus of very young stars (<100\,Myr) and also distinct from the Hyades. We adopt an age of 4\,Gyr.\\
Owing to the equatorial coordinates of the AE For system, our two SPHERE epochs collected a very poor sky rotation. The first epoch achieves an increment in parallactic angle of only 1.4$^{\circ}$, which does not allow us to get any information closer than 0.7\,$\arcsec$ to the star. Our second epoch obtains $\sim$3 degrees of sky rotation and poor contrast at close separations. We do not find any candidate companion, but at large separations ($>$50\,AU) we can rule out objects more massive than $\sim$10\,$M\rm_{Jup}$.\\

\noindent
\textbf{HIP 14807 = BD+21 418B:} Visual binary in the AB Doradus moving group. \citet{evans2012} resolved for the first time a secondary companion at Keck with aperture-masking interferometry in the CO filter, and estimated a mass of 0.52\,$\pm$\, 0.09\,\(M_\odot\) via evolutionary models for an age of 110\,$\pm$\, 40\,Myr. The authors claim that the mass value includes the uncertainty in age and distance, which is compatible with our AB Dor age estimate of 149\,Myr taken from \citet{bell2015}. We then adopt their estimated masses for the primary and secondary.\\

\noindent
\textbf{HIP 15197 = $\zeta$ Eri:} Early-type SB1 hosting a circumbinary debris disk identified by IRAS/MIPS satellite \citep{rhee2007}. This SB1 was later resolved in the far-IR by $Herschel$ at about 96\,AU \citep{booth2013}. The authors also suggested that a broad ring or a second inner ring would improve the fit. This binary shows a period of 18 days and an eccentricity of 0.14 \citep{pourbaix2004,abt2005}. Kinematics are compatible with a star of a few hundred Myrs. Different isochronal works point to somewhat diverese ages; \citet{rhee2007} and \citet{derosa2014} obtain an age estimation of about 400\,Myr, while \citet{vican2012} adopt an age of 800\,Myr. We adopt the isochrone age from \citet{vican2012}.\\  

\noindent
\textbf{HIP 16853 = HD 22705 :} Tucana association member with a likely maximum separation of 18\,mas; see SPOTS I and SPOTS II. A star is found in our NaCo observation (see Table \ref{binaries}), forming a triple system. This object had also been discovered with NICI \citep{tokovinin2012} almost contemporaneously with our observations, and was reported as a companion by \citet{galicher2016}.\\ 

\noindent
\textbf{HIP 19591 = V1136~Tau = HD 284163:} Triple system in the Hyades. The inner $\sim$2-day orbit inner pair is accompanied by a close NaCo-resolved companion in a 40\,yr orbit. See SPOTS I. We resolve the outer companion; see Table \ref{binaries}.\\ 

\noindent
\textbf{TYC 5907-1244-1:}  SB2  surrounded by a newly discovered candidate circumbinary debris disk at about 50\,AU and T $\sim$ 30\,K \citep{moor2016}. See SPOTS I. \\

\noindent
\textbf{HIP 25486 = AF~Lep:}  $\beta$ Pic SB2 object, see SPOTS I.\\

\noindent
\textbf{HIP 25709 = HD 36329:} SB2 in the Columba association, see SPOTS I. \\

\noindent
\textbf{HIP 27134 = XZ~Pic:} Short-period SB1, see SPOTS I. \\

\noindent
\textbf{HIP 31681 = Alhena = $\gamma$~Gem:} See detailed study and description in SPOTS I.\\

\noindent
\textbf{TYC 8104 0991 1: } SB3, see SPOTS I. We resolve the outer third stellar companion for the first time at 133\,mas and estimate a mass of 0.78\,\(M_\odot\), see Table \ref{binaries}. \\

\noindent
\textbf{HIP 32104 = 26 Gem = HD 48097 = HR 2466:} Spectroscopic and astrometric binary member of Columba association. See SPOTS I.\\

\noindent
\textbf{EM Cha = RECX7:} SB2 in the $\eta$ Cha cluster, see SPOTS I.\\

\noindent
\textbf{TYC 8569 3597 1:} SB2 member of the Carina moving group, see SPOTS I.  This target was observed both with NaCo and SPHERE in a crowded field with CCs that do not agree well with common proper motion nor background movement. We will consider them as background objects given the astrometric uncertainty of NaCo. Moreover, adding a common proper motion shift to all the candidates puts the seven of these in a clear background motion. \\

\noindent
\textbf{TYC 9399-2452-1 = HD 81485B:} SB2 that forms part of a quadruple system at about 500\,AU of HD 81485A, a close visual binary with masses of about 0.8\,\(M_\odot\) \citep{tokovinin2010}. See SPOTS I. An initial NaCo epoch detected a hypothetical 10\,$M\rm_{Jup}$ planet at about 4.25\,$\arcsec$. SPHERE redetected this candidate three years later, which confirmed it as a background object.  \\

\noindent
\textbf{HIP 46509 = $\tau$ Hya:} System formed by a SB1 with a F6V spectral type primary and a wide third K-type main sequence star candidate at about 66\,$\arcsec$ \citep{montesinos2016}. This distant component has a low probability of constant RV from \citet{nordstrom2004} and \citet{casagrande2011} monitoring. It could however be a SB itself, forming a quadrupole system with the HIP 46509 inner binary. The period and eccentricity of the inner SB1 is revised by \citet{hal2012}. Several methods have been used to date this system, such as chromospheric activity and X-ray emission, estimating ages of 1.44 and 0.86\,Gyr, respectively \citep{vican2012}. Also, two ages from isochrone fitting of 1.8 and 2.5 Gyr are derived by \citet{holmberg2009} and \citet{david2015}, respectively. Kinematics parameters also point to an age older than 1\,Gyr. We adopt an age of 2\,Gyr.\\
No CCs are found in any of the two SPHERE epochs, but we resolved the spectroscopic binary for the first time at $\sim$0.35\,$\arcsec$; see Table \ref{binaries}. From a visual photometric analysis, we derive a  magnitude difference of $\Delta$H2$\sim$4.7, which corresponds to a mass for the companion of $\sim$0.35\,\(M_\odot\) for an age of 2\,Gyr.  \\

\noindent
\textbf{HIP 46637 = GS~Leo:} Close binary in a 3.86 day period with a wide tertiary component at $\sim$14\,\arcsec (or 508\,AU). See \citet{desidera2015}. As explained in SPOTS I, for this target we acquired a follow-up observation with the Subaru IRCS camera, which discarded the presence of a candidate near IWA previously identified by NaCo.\\

\noindent
\textbf{HIP 47760 = HD 84323:} SBI, see SPOTS I. \\

\noindent
\textbf{TYC 6604 0118 1:} Very close SB2, see SPOTS I.\\

\noindent
\textbf{HIP 49669 = $\alpha$ Leo = Regulus:} Quadruple system with a close stellar pair at 175\,\arcsec, see SPOTS II.\\

\noindent
\textbf{HIP 49809 = HD 88215:} SB1 hosting a debris disk, see SPOTS II. \\


\noindent
\textbf{HIP 56960 = HD 101472:}  Very close X-ray source SB1 with a doubtful wide companion of 0.33\,\(M_\odot\) at about 80\,$\arcsec$, see \citet{tokovinin2014}. A more recent proper motion analysis appears to reject this association. The binary system has been studied with high-contrast imaging at the Palomar telescope, without finding any other companions \citep{metchev2009}. The age of this field binary varies depending on the dating method. \citet{metchev2009} announced an age of 250\,Myr from the strength of chromospheric activity emission lines, while \citet{weise2010} inferred a younger system of 90\,Myr via lithium indications. As the system is probably tidally locked, we rely only on lithium and kinematics, adopting an age of 300\,Myr. The mass of the primary was estimated from the absolute $V$ magnitude of the binary component. The mass of the secondary component is derived as a minimum mass inferred from the mass function and the mass of the primary component \citep{tokovinin2014}.\\

\noindent
\textbf{HIP 57363 = $\lambda$ Muscae: } Astrometric binary in a close 1.24\,yr-period orbit with a 0.3 eccentricity \citep{hipp} and an early-type primary. \citet{Hartkopf2012} observed the system with speckle interferometry at the SOAR telescope, but did not detect any other companion. Age and mass taken from the \citet{david2015} isochronal models. We discover a low-mass stellar companion in two epochs: May 2016, ($\Delta$Sep($\arcsec$), PA($\deg$)) = (0.158\,$\pm$\, 0.003, 6.5\,$\pm$\, 1.0), and June 2016, ($\Delta$Sep($\arcsec$), PA($\deg$)) = (0.153\,$\pm$\, 0.004, 2.7\,$\pm$\, 1.8). See
section \ref{lambdamus} for more information.\\

\noindent
\textbf{TYC 9412-1370-1:}  X-ray SB1 with an orbital period of 614 days \citep[see][]{guenther2007}, where the primary is a K-type star with a rotational period of $\sim$3\,days \citep{messina2011}. \citet{guenther2007}  considered this binary to be a WTTS in the Chameleon star-forming region, but later studies assigned it Argus membership \citep[e.g.,][]{dasilva2009,desilva2013}. The Li EW from the SACY project \citep{elliot2016} is a much better match to Argus age. 
We adopt Argus membership. \\ 
Our two SPHERE observations show a $\geq$5\,$\sigma$ background object at $\sim$1.7\,\arcsec. Other three background objects are detected at a lower S/N.\\

\noindent
\textbf{HIP 63742 = PX Vir:}  AB Dor member, see SPOTS II.\\


\noindent
\textbf{HIP 69781 = V* V636 Cen:} Very close solar-type eclipsing and SB2 binary. Both the secondary and primary show signs of high chromospheric activity and starspots, especially the secondary, as they are probably tidally locked in a 4 day period. An old age of 3\,Gyr is derived from evolutionary models \citep{clausen2009, fernandes2012}, but a few arguments suggest the contrary. For instance, a non-zero eccentricity is difficult to explain at such old age, unless there is an unseen outer companion. The lithium EW from \citet{torres2006} would indicate an age of 100--200\,Myr, and the kinematics are in agreement with the young disk. We decided to adopt an age of 250\,Myr, which could be compatible with lithium, kinematics and eccentricity. Masses are adopted from \citet{clausen2009} at 0.5$\%$ precision. 

Crowded field with no comoving companions, see Figure \ref{im}.\\

\noindent
\textbf{HIP 74049  = HS~Lup :} SB2 with a 17.83 day period and q = 0.983 \citep{tokovinin2014}. From an estimation of the primary mass from the visual absolute magnitude of the binary system, we use the mass ratio to obtain the mass of the secondary. See also SPOTS I. Two wide candidates are detected by NaCo, but only one of these is recovered  later within SPHERE's FoV at $\sim$5\,\arcsec, which was determined to follow background motion.\\

\noindent
\textbf{1RXS J153557.0-232417 = GSC 06764-01305:} Close visual binary, see SPOTS II.\\
Even though the observation collected an extremely poor sky rotation, SPHERE detects two CCs at large separation. No follow-up epoch 
was acquired.\\

\noindent
\textbf{HIP 76629 = HD 133822:}  $\beta$ Pic moving group binary with an additional wide M5-companion at $\sim$10\,$\deg$. \citet{nielsen2016} has recently resolved the inner binary with GPI; see their full analysis. Together with archival NaCo imaging, RV, and astrometric measurements, they derived individual dynamical masses and a full orbital solution that we adopt here. See also SPOTS I.\\
Very crowded field, no comoving companion found.\\

\noindent
\textbf{HIP 77911 = HD 142315:}  US triple system with a wide companion at $\sim$8\,$\arcsec$ discovered by an ADONIS adaptive optics Sco-Cen survey \citep{kouwe2005}. The SEEDS High-Contrast Imaging Survey \citep{uyama2017} also observed the inner system with HiCIAO, finding some CCs at $>$4\,$\arcsec$. We adopt the orbital parameters for the inner SB1 derived by \citet{levato1987}. Moreover, this target hosts an unresolved debris disk at about 40\,AU \citep{jang2015}. We use the mass function from \citet{levato1987} to derive a minimum mass of the companion from our photometric mass estimation of the primary. \\ 
Although we only count with one epoch for HIP 77911, we use HiCIAO's previous observation to do the astrometric analysis of the three CCs detected by SPHERE between 4 and 5\,$\arcsec$. The nature
of the three objects is not clear. See Section \ref{sec:77911}.\\

\noindent
\textbf{HIP 78416 = HD 143215:}  SB2 orbited by a wide $\sim$1.15\,\(M_\odot\) companion at 6.55\,$\arcsec$ (550\,AU); see SPOTS I. Age and kinematics compatible with UCL membership.\\
The star is in a crowded field with no comoving companions\\

\noindent
\textbf{RX J1601.9-2008 = BD-19 4288:}  X-ray source located in the US subregion \citep{kohler2000}. \citet{rizzuto2016} monitored this binary with adaptive optics via sparse aperture masking during 8 years, finding a semimajor axis of 36\,mas. We adopt their orbital solution and masses. Their age estimate is consistent with the reported value by \citet{pecaut2012} for US, which we adopt.\\
SPHERE detects two background candidates beyond 3\,$\arcsec$.\\

\noindent
\textbf{ScoPMS027 = V1156 Sco:} Close visual binary; see SPOTS II. Two $\sim$4$M\rm_{Jup}$ candidates at very wide separation. Epochs are one month apart, too close for common proper motion analysis. Given the distance to the central binary they probably move with the background.\\

\noindent
\textbf{TYC 6209-735-1:} Discovered as a SB1 by \citet{guenther2007}. This binary has been resolved from sparse aperture mask observations by \citet{rizzuto2016} and \citet{kraus2008}. We adopt their orbital solution. Both the single-lined binarity information and the visual orbit are put together by \citet{rizzuto2016} to produce posteriors that constrain the stellar masses that we adopt. The system parallax is consistent with both US and UCL membership, but the authors estimate isochronal ages that are closer to UCL. We then assume UCL membership.\\
Our SPHERE observation only reveals a $\sim$7\,$M\rm_{Jup}$ candidate and a likely faint background galaxy at very large separation. No follow-up epoch available for this target.\\

\noindent
\textbf{ScoPMS044 = V1000 Sco = Wa Oph 1:} US single-lined spectroscopic binary, whose period and eccentricity was estimated by  \citet{mathieu1989}. The secondary has recently been resolved by \citet{anthonioz2015} at a projected separation of 4.33\,mas, although no full orbital solution has been obtained yet. We adopt a distance of 145\,pc to US \citep{dezeeuw1999} and estimate a mass for the primary accordingly.\\

\noindent
\textbf{ScoPMS048 = V1001 Sco:} Triple system in the US Sco-Cen subregion. A wide 0.4\,\(M_\odot\) companion has been resolved at a distance of $\sim$3\,arcsec (440\,AU) with a mass ratio of $q = 0.78$ with respect to the inner SB1 \citep{kraus2009}. We adopt the SB1 period and eccentricity derived by \citet{mathieu1989}, and the mass of the primary from \citet{kraus2008}. We detect the wide stellar companion (see Table \ref{binaries}) but no candidate substellar companions.\\

\noindent
\textbf{TYC 6213-306-1  = 1RXS J161318.0-221251:} Equal-mass SB2, see SPOTS II. This target was observed with the SPHERE side of the SPOTS survey and revealed the presence of two 5--6\,$M\rm_{Jup}$ candidates at $\sim$3.5\,$\arcsec$. Although we could not obtain a follow-up for this target, 
these two candidates had previously been revealed in 2008 by \citet{lafreniere2014} with the Gemini North telescope. Using the coordinates provided by that work, we confirm that both candidates follow the background trajectory.\\

\noindent
\textbf{HD 147808 = GSC 06794-156:} Similar-mass binary recently resolved from sparse aperture mask observations by \citet{rizzuto2016}. We adopt their orbital solution and masses. Evolutionary model ages and the system parallax agree well with Sco-Cen US  membership. A candidate companion was found at 6\,$\arcsec$ by \citet{ireland2011}, but later confirmed as background star by \citet{kraus2014}.\\
Our only SPHERE observation acquires only 6\,$\deg$ of sky rotation, but we detect a $\sim$3\,$M\rm_{Jup}$ candidate at about 2\,$\arcsec$ (see Table \ref{unknowns}). This candidate will have to be checked for common proper motion.\\

\noindent
\textbf{HIP 80686 = $\zeta$ TrA}: SB1 associated with the Ursa Major moving group. This target has been observed with coronagraphic observations with NACO by \citet{ammler2016}, finding several CCs at large separations ($>$\,5\,$\arcsec$). The spectroscopic orbit was derived by \citet{skuljan2004}, who estimated that the secondary must be an M-type star (or $<$0.45\,\(M_\odot\)), given the absence of spectroscopic signature, at high enough inclination to produce the RV signal. From the mass function they derived a lower limit of 0.094\,\(M_\odot\) , taking the mass of the primary as 1.12\,\(M_\odot\) from evolutionary models. We adopt a mass for the secondary of 0.2\,\(M_\odot\) \\

\noindent
\textbf{ROX 33 =  EM* SR 20:}  Young accreting system in the $\rho$ Ophiuchi star-forming region with close companions separated by $\sim$5--10\,AU. The system presents a likely circumstellar disk  truncated beyond 0.39\,AU from the primary, as seen from the lack of spectro-astrometic signature in H$\alpha$ \citep{mcclure2008}. It has recently been resolved with Keck aperture masking observations at a separation of 50\,mas \citep{cheetham2015} and mass ratio $q = 0.25$. Together with previous speckle imaging observations \citep[][]{ghez1993,ghez1995}, an orbital solution could be obtained, but this has not been done yet. As there is no trigonometric parallax for this system, we assume 120\,pc as the distance to the $\rho$ Ophiuchi cloud \citep{loinard2008}. We adopt the masses from \citet{cheetham2015}.  \\

\noindent
\textbf{ROXs 43A:} Pre-main sequence SB1 in the  $\rho$ Ophiuchi cloud with an orbital period of about 89\,days \citep{mathieu1989} and possibly a circumbinary disk \citep{jensen1997}. This SB1 (A) is in a hierarchical system that includes an outer binary companion at $\sim$4.5\,\arcsec (C) with components separated by 16\,mas \citep{simon1995}. Moreover, \citet{correia2006} detected a low-mass companion (B) to the A binary at about 0.3\,\arcsec with NACO direct imaging. Using the new Gaia-DR2 distance and assuming $\rho$ Ophiuchi cloud age, we derive the masses of the A, B, and C components. \\
Very bad observation with almost no sky rotation. The stellar companion at $\sim$0.3\,\arcsec (B) and the wide outer binary (C) can be detected in the unsaturated frames at positions with respect to A of 
($\Delta$Sep($\arcsec$), PA($\deg$)) = (0.2953\,$\pm$\, 0.0013, 156.8\,$\pm$\, 0.3) and ($\Delta$Sep($\arcsec$), PA($\deg$))= (4.484\,$\pm$\, 0.003, 11.45\,$\pm$\, 0.14), respectively. \\

\noindent
\textbf{HIP 82747 = AK Sco:} SB2 binary, see \citet{janson2016}. The binary has also been resolved and thus the full orbital solution and masses were obtained by \citet{anthonioz2015}.\\
Our two SPHERE observations revealed the presence of a circumbinary disk, which led to the \citet{janson2016} publication. The surrounding sky field to the binary is crowded with CCs,
as expected from its galactic position, but none of these were found to be comoving.\\

\noindent
\textbf{HIP 84586 = HD 155555:} Triple system formed by a close SB2 and a very wide tertiary companion, see SPOTS II. A $\sim$5\,$M\rm_{Jup}$ candidate is visible in our first SPHERE epoch at the edge of the IRDIS FoV, but lays outside in the second epoch.
Judging by its large distance to the binary star, it is very unlikely that this is a comoving planet-like companion.   \\

\noindent
\textbf{HIP 88481 = HD 165045:} Similar-mass double-lined and visual binary. It has been resolved by \citet{tokovinin2017} with speckle interferometry. Combining resolved images and RV measurements, they find a low-inclination orbital solution and dynamical masses that we use here. Lithium and activity indications \citep{isaacson2010,cutispoto2003} suggest an age similar to the Hyades. We adopt 600\,Myr for this system.  \\
Two SPHERE epochs showed a candidate at about 5$\arcsec$, which moved with the background.\\

\noindent
\textbf{HIP 94050 = HD 177996:} Short-period SB2, see SPOTS II. \\
A potential candidate at $\sim$2.3\,$\arcsec$ was detected, but showed background motion. \\

\noindent
\textbf{HIP 98704 = HD 188480:} Flagged as an astrometric binary by \citet{makarov2005} and as a spectroscopic binary by \citet{frankowski2007}. No orbital solution is available. The astrometric binarity seems to argue against tidal locking if no inner additional components exist, as suggested by the relatively small RV dispersion in \citet{nordstrom2004}. This would imply a period longer than 1\,yr, in which case the RHK and X-ray emission yield an age of about 250\,Myr \citep{henry1996}. On the other hand the kinematics, based on two RV epochs \citep{nordstrom2004}, point to outside the kinematic space of young stars. We adopt an age of 250\,Myr, but with an upper limit of several Gyrs in the case of tidal locking. \\
Three candidates encountered with background motion.\\ 

\noindent
\textbf{HIP 100751 = $\alpha$ Pav:} Close spectroscopic binary, member of Tucana. See SPOTS II. \\

\noindent
\textbf{HIP 101422 = HD 195289:}  SB1 indicated as such given its high RV scatter, which seems to be a clear indicator of binarity \citep{kharchenko2007}.  It was detected as an X-ray source \citep{schwope2000}. The large lithium EW from \citet{torres2006} would imply a very young system of $<$100\,Myr, but isochronal fitting from \citep{casagrande2011} favors an age of 3\,Gyr. If the binary is tidally locked, the old isochronal age might explain the chromospheric activity, but not the lithium signature. Kinematics do not allow to rule out a young age. We decide to rely on the lithium signature, and adopt an age of 100\,Myr. No orbital solution is available.\\ 

\noindent
\textbf{HIP 104043 = $\alpha$ Oct:}  This spectroscopic and eclipsing binary is composed of two components in a 9-day orbit. An infrared excess has been detected, which probably indicates that this system hosts a circumbinary debris disk at about 10\,AU \citep{trilling2007}. The UVW space velocity of this system is well outside the kinematic space of young stars, which allows us to infer an age older than 1\,Gyr. This is consistent with the isochronal age reported by \citet{casagrande2011} of 1.1\,Gyr. We adopt the age and mass from \citet{casagrande2011}. \\

\noindent
\textbf{HIP 105404 = BS~Ind:} Triple system formed by a SB1 composed of a K0V primary star that is orbited every 3.3\,yr by a pair of eclipsing late-K or early-M stars in a 0.43-day period, according to \citet{guenther2005}. This work  adopted a Hipparcos distance of 46\,pc to estimate a mass of the primary of 0.8\,\(M_\odot\). This distance is similar to the recent 52.7\,pc Gaia-DR2, which we adopt.
Now, from here we use the SB1-derived mass function by \citet{guenther2005} to estimate a minimum total mass for the eclipsing pair of 0.9\,\(M_\odot\). Assuming they are both equal mass, we consider a mass of 0.45\,\(M_\odot\) each, and include it in Table \ref{tab:mastertable}. The very strong lithium line of this system indicates a very young system. We adopt an age of 30\,Myr.\\
A very poor SPHERE follow-up observation allows us to rule out common proper motion of two candidates detected by NaCo.\\
 
\noindent
\textbf{HIP 105860 = IK Peg  = HD 204188 : } SB1 with a massive white dwarf secondary, see SPOTS I.\\
Two candidates at large separations were detected by NaCo. Only one of these was within the FoV of SPHERE, which showed a clear background motion. An additional 7\,$M\rm_{Jup}$
potential candidate near IWA was flagged in SPOTS I with NaCo, but could not be recovered with SPHERE. We then assume it to be a residual speckle feature.\\  

\noindent
\textbf{HIP 107556 = $\delta$ Cap :} Spectroscopic and eclipsing binary. See SPOTS I.\\

\noindent
\textbf{HIP 109110 = NT Aqr:} X-ray source and variable star form an astrometric, SB1, and visual binary. The spectroscopic observations deliver a preliminary period of 13.1\,yr and expected semimajor axis of 0.19\,\arcsec \citep{tokovinin2014}. The system appears to be resolved by \citet{tokovinin2013} with NICI adaptive optics at about 0.08\,$\arcsec$, which may give a period estimate of  $\sim$4\,yr. However, the secondary shows up below the formal NICI detection limit and had not been resolved in previous studies \citep[e.g.,][]{metchev2009}. We adopt the spectroscopic period, in which case the binary should not be tidally locked and the derived age of 0.7\,Gyr from rotation by
 \citet{baumann2010} might be a good approximation. \citet{casagrande2011} estimated an age of 10\,Gyr from isochronal fitting, but with a distance estimation that is inconsistent with Gaia-DR2, and without considering the effect of binarity. We adopt an age of 1\,Gyr for this system.\\ 

\noindent
\textbf{HIP 109901  = CS~Gru :}  SB1, see SPOTS I. \\

\noindent
\textbf{TYC 6386 0896 1:} SB2 binary with lithium content similar to Pleiades stars, probably tidally locked.  See SPOTS I. \\

\noindent
\textbf{HIP 113860 = pi. PsA:} Astrometric and spectroscopic binary with an IR excess \citep{chen2014}. We adopt the orbital elements derived by \citet{bopp1970}. Kinematics exclude very small ages and, assuming that the X-ray emission of the system comes from the early type secondary,  its luminosity falls just below the median value of the Pleiades (125\,Myr) and well above the Hyades (625\,Myr). This is in line with the isochronal age of 175\,Myr from \citet{david2015}, which we adopt. \\

 \begin{figure*}
  \resizebox{\hsize}{!}{\includegraphics{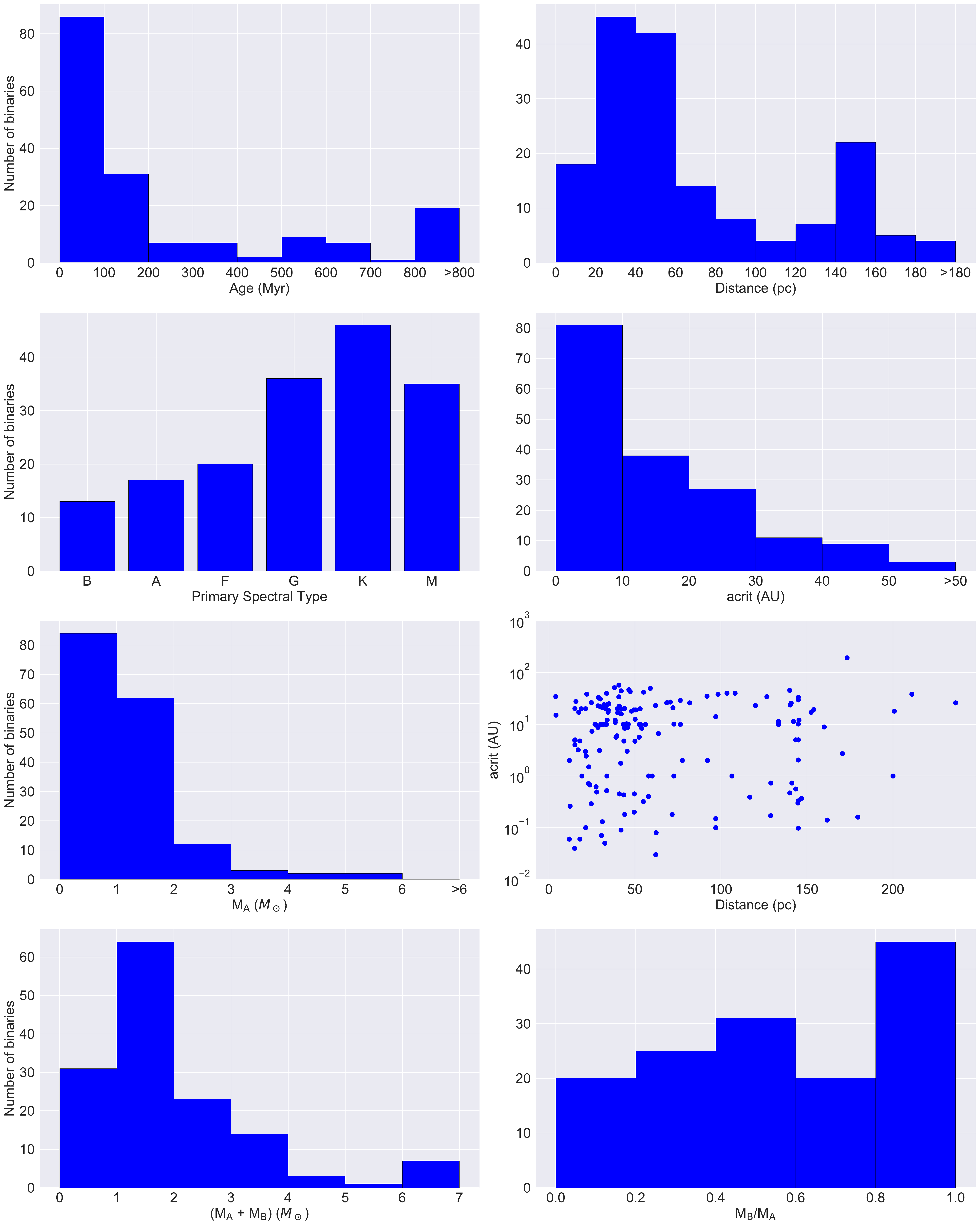}}
  \caption{Same as Figure \ref{histogram}, but for the targets used in the SPOTS II + III statistical analysis.}
  \label{histogram_IIandIII}
\end{figure*}

\clearpage

\begin{table*}
\caption{Best contrast achieved in the SPOTS observations at certain separations from the binary star  }
\centering
\begin{tabular}{l cccccc}
\hline \hline
Target & 0.1\,$\arcsec$ & 0.25\,$\arcsec$ & 0.5\,$\arcsec$ & 1\,$\arcsec$ & 2\,$\arcsec$ & 3\,$\arcsec$ \\
\hline
HIP 7601 & 4.19e-02 & 1.20e-03 & 1.08e-04 & 1.84e-05 & 3.63e-06 & 3.28e-06 \\
HIP 9892 & -- & -- & 1.69e-04 & 2.12e-05 & 3.79e-06 & 2.40e-06 \\
HIP 12225 & 4.09e-03 & 1.82e-04 & 7.70e-05 & 2.79e-04 & 6.50e-05 & 6.57e-05 \\
HIP 12545 & -- & 9.65e-04 & 6.47e-05 & 6.48e-06 & 1.66e-06 & 1.36e-06 \\
HIP 12716 & -- & 1.09e-03 & 1.14e-04 & 8.71e-06 & 2.21e-06 & 1.97e-06 \\
HIP 14007 & 8.80e-04 & 1.18e-05 & 4.55e-06 & 3.17e-06 & 1.32e-06 & 1.17e-06 \\
HIP 14568 & 4.00e-04 & 5.36e-05 & 3.69e-05 & 3.66e-05 & 2.34e-06 & 1.54e-06 \\
HIP 14807 & 1.87e-03 & 5.13e-05 & 9.75e-06 & 8.14e-06 & 3.00e-06 & 2.64e-06 \\
HIP 15197 & 2.87e-03 & 1.26e-05 & 4.34e-06 & 6.63e-06 & 1.01e-06 & 8.53e-07 \\
HIP 16853 & -- & 6.34e-04 & 6.81e-05 & 6.79e-06 & 1.82e-06 & 1.31e-06 \\
HIP 19591 & -- & -- & 1.00e-03 & 2.37e-05 & 4.28e-06 & 3.28e-06 \\
TYC 5907-1244-1 & -- & 5.64e-04 & 5.45e-05 & 6.15e-06 & 1.77e-06 & 1.34e-06 \\
HIP 25486 & -- & 1.40e-03 & 1.72e-04 & 2.01e-05 & 3.39e-06 & 2.23e-06 \\
HIP 25709 & -- & 1.00e-02 & 5.76e-05 & 3.99e-06 & 1.22e-06 & 1.02e-06 \\
HIP 27134 & -- & 1.97e-03 & 3.35e-04 & 3.23e-05 & 7.35e-06 & 5.17e-06 \\
HIP 31681 & -- & -- & 1.34e-04 & 5.15e-06 & 6.57e-07 & 2.65e-07 \\
TYC 8104-0991-1 & -- & 2.45e-03 & 5.00e-04 & 2.39e-05 & 6.21e-06 & 5.18e-06 \\
HIP 32104 & -- & 1.04e-03 & 1.28e-04 & 9.06e-06 & 1.61e-06 & 9.49e-07 \\
Em Cha & -- & 1.90e-04 & 5.63e-04 & 4.82e-05 & 1.21e-05 & 1.02e-05 \\
TYC 8569-3597-1 & -- & 1.01e-04 & 1.71e-05 & 9.86e-06 & 3.31e-06 & 2.34e-06 \\
TYC 9399-2452-1 & -- & 4.26e-05 & 1.07e-05 & 6.56e-06 & 1.47e-06 & 1.20e-06 \\
HIP 46509 & 1.72e-05 & 1.18e-05 & 3.32e-06 & 3.97e-06 & 1.34e-06 & 1.30e-06 \\
HIP 46637 & -- & 3.52e-04 & 3.98e-05 & 5.89e-06 & 2.02e-06 & 1.38e-06 \\
HIP 47760 & -- & 5.07e-04 & 4.51e-05 & 5.01e-06 & 1.52e-06 & 1.19e-06 \\
TYC 6604-0118-1 & -- & -- & 1.40e-03 & 9.63e-05 & 8.96e-06 & 4.41e-06 \\
HIP 49669 & 1.03e-03 & 1.51e-05 & 8.68e-06 & 1.21e-05 & 1.89e-06 & 1.55e-06 \\
HIP 49809 & 1.54e-04 & 2.47e-05 & 8.05e-06 & 7.09e-06 & 1.13e-06 & 9.98e-07 \\
HIP 56960 & 1.57e-03 & 8.35e-05 & 1.13e-05 & 9.39e-06 & 1.92e-06 & 1.60e-06 \\
HIP 57363 & 1.26e-04 & 1.51e-05 & 7.34e-06 & 6.61e-06 & 1.88e-06 & 1.80e-06 \\
TYC 9412-1370-1 & 4.81e-03 & 3.79e-05 & 1.46e-05 & 5.85e-06 & 2.74e-06 & 2.57e-06 \\
HIP 63742 & 2.00e-04 & 1.68e-05 & 7.65e-06 & 2.62e-06 & 7.29e-07 & 6.09e-07 \\
HIP 69781 & 4.64e-04 & 2.58e-05 & 4.45e-06 & 4.97e-06 & 1.45e-06 & 1.35e-06 \\
HIP 74049 & 6.71e-05 & 9.22e-06 & 3.04e-06 & 4.03e-06 & 1.26e-06 & 1.14e-06 \\
1RXS J1535570 232417 & 1.76e-02 & 5.04e-05 & 9.42e-05 & 2.42e-05 & 9.15e-06 & 8.15e-06 \\
HIP 76629 & 3.33e-05 & 1.30e-05 & 6.82e-06 & 3.31e-06 & 9.41e-07 & 7.73e-07 \\
HIP 77911 & 7.95e-05 & 2.19e-05 & 9.35e-06 & 5.74e-06 & 1.30e-06 & 1.09e-06 \\
HIP 78416 & 4.59e-06 & 2.43e-05 & 6.98e-06 & 3.68e-06 & 1.69e-06 & 1.54e-06 \\
RXJ1601.9-2008 & 3.34e-06 & 1.35e-05 & 4.61e-06 & 2.58e-06 & 1.27e-06 & 1.26e-06 \\
ScoPMS027 & 5.58e-05 & 5.30e-05 & 6.41e-06 & 5.28e-06 & 2.74e-06 & 2.57e-06 \\
TYC 6209-735-1 & 2.39e-04 & 2.64e-05 & 1.41e-05 & 5.43e-06 & 2.72e-06 & 2.48e-06 \\
ScoPMS044 & 1.04e-03 & 2.83e-05 & 2.01e-05 & 4.05e-06 & 1.92e-06 & 1.75e-06 \\
ScoPMS048 & 1.76e-03 & 1.82e-04 & 6.77e-05 & 5.30e-06 & 3.04e-06 & 3.02e-06 \\
TYC 6213-306-1 & 8.30e-03 & 9.25e-04 & 5.53e-05 & 6.93e-06 & 2.10e-06 & 1.65e-06 \\
HD 147808 & 6.25e-03 & 7.76e-05 & 2.72e-05 & 7.71e-06 & 2.04e-06 & 1.63e-06 \\
HIP 80686 & 5.49e-04 & 9.29e-05 & 5.62e-06 & 7.92e-06 & 1.55e-06 & 1.45e-06 \\
ROX 33 & 2.20e-05 & 6.99e-06 & 6.62e-06 & 2.98e-05 & 8.68e-06 & 4.50e-06 \\
ROXs 43A & 6.98e-04 & 2.71e-04 & 3.87e-05 & -- & 6.87e-06 & 4.34e-06 \\
HIP 82747 & 6.67e-04 & 2.99e-05 & 1.03e-05 & 3.03e-06 & 1.66e-06 & 1.21e-06 \\
HIP 84586 & 3.05e-04 & 1.28e-05 & 5.76e-06 & 4.45e-06 & 1.25e-06 & 1.21e-06 \\
HIP 88481 & 2.32e-04 & 1.02e-05 & 4.30e-06 & 3.72e-06 & 1.31e-06 & 1.34e-06 \\
HIP 94050 & 2.80e-04 & 2.41e-05 & 6.22e-06 & 5.04e-06 & 1.19e-06 & 1.03e-06 \\
HIP 98704 & 1.00e-04 & 2.90e-05 & 1.29e-05 & 6.14e-06 & 1.84e-06 & 1.52e-06 \\
HIP 100751 & 8.35e-06 & 3.69e-05 & 9.11e-06 & 3.65e-06 & 3.80e-07 & 2.56e-07 \\
HIP 101422 & 7.89e-05 & 6.50e-06 & 4.09e-06 & 6.98e-06 & 1.63e-06 & 1.54e-06 \\
HIP 104043 & 7.02e-05 & 2.55e-05 & 1.87e-06 & 5.72e-07 & 1.37e-07 & 1.27e-07 \\
HIP 105404 & 1.00e-02 & 1.90e-03 & 1.79e-04 & 1.47e-05 & 3.18e-06 & 2.14e-06 \\
HIP 105860 & 6.04e-05 & 1.79e-05 & 6.70e-06 & 5.94e-06 & 1.70e-06 & 1.57e-06 \\
HIP 107556 & -- & 7.96e-04 & 7.58e-05 & 8.11e-06 & 1.29e-06 & 4.99e-07 \\
HIP 109110 & 2.32e-04 & 9.86e-06 & 5.33e-06 & 5.55e-06 & 1.24e-06 & 1.10e-06 \\
HIP 109901 & -- & 2.24e-03 & 1.52e-04 & 1.59e-05 & 6.00e-06 & 4.63e-06 \\
TYC 6386-0896-1 & -- & -- & 2.41e-03 & 4.44e-05 & 6.16e-06 & 3.66e-06 \\
HIP 113860 & 7.87e-04 & 2.23e-05 & 9.49e-06 & 1.48e-05 & 1.96e-06 & 1.69e-06 \\
\hline
\end{tabular}

\end{table*}

\clearpage

\begin{table*}
\caption{Imaged point sources (not including known higher order systems and crowded fields)}
\centering
\scriptsize{
\begin{tabular}{l c c c c c c c c}
\hline \hline
Target & CC & $\Delta$\,Sep ($\arcsec$) & $\Delta$\,PA ($\deg$) & $\Delta$H & $\Delta$H2 &  $\Delta$H3 & Epoch & Status \\
\hline
HIP 7601 & a & 2.826\,$\pm$\,0.002 & 23.10\,$\pm$\,0.14 & -- & 7.76\,$\pm$\,0.16 & 7.71\,$\pm$\,0.11 & 2015-08-25 & Unknown \\
HIP 12225 & a & 5.31\,$\pm$\,0.02 & 238.0\,$\pm$\,0.3 & -- & 9.4\,$\pm$\,0.5 & 10.1\,$\pm$\,0.5 & 2015-08-28 & Bckg \\
HIP 31681 & a & 1.833\,$\pm$\,0.003 & 162.94\,$\pm$\,0.06  & 12.5\,$\pm$\,0.2  & --  & -- & 2011-12-22 &  Bckg\\
HIP 31681 & b & 2.493\,$\pm$\,0.005 & 150.07\,$\pm$\,0.09 &  14.28\,$\pm$\,0.17  & -- & -- & 2011-12-22 &  Bckg\\
HIP 31681 & a & 1.791\,$\pm$\,0.008 & 163.6\,$\pm$\,0.3  &  12.0\,$\pm$\,0.19  &  -- & -- & 2013-03-01 & Bckg\\
HIP 31681 & b & 2.447\,$\pm$\,0.012 & 150.4\,$\pm$\,0.3 & 13.94\,$\pm$\,0.15 & & & 2013-03-01 & Bckg \\
Em Cha    & a &  5.031\,$\pm$\,0.008    & 348.13\,$\pm$\,0.06     & 6.1\,$\pm$\,0.2 &  -- & --  & 2011-12-14 & Bckg\\
TYC 9399-2452-1 & a & 4.28\,$\pm$\,0.02 & 182.4\,$\pm$\,0.3 & 12.99\,$\pm$\,0.14    & -- & --   & 2013-03-03 & Bckg\\
TYC 9399-2452-1 & a & 4.485\,$\pm$\,0.005 & 179.07\,$\pm$\,0.14 & -- & 13.22\,$\pm$\,0.11 & 13.35\,$\pm$\,0.11 & 2016-04-03 & Bckg\\
HIP 57363 & a & 0.158\,$\pm$\,0.003 & 6.5\,$\pm$\,1.0 & -- & 8.4\,$\pm$\,1.1 & 8.3\,$\pm$\,1.1 & 2016-05-07 & Comoving \\
HIP 57363 & a & 0.153\,$\pm$\,0.004 & 2.7\,$\pm$\,1.8 & -- & 8.2\,$\pm$\,0.9 & 8.0\,$\pm$\,0.9 & 2016-06-03 & Comoving \\
TYC 9412-1370-1 & a & 1.727\,$\pm$\,0.003 & 58.95\,$\pm$\,0.17   & -- & 11.87\,$\pm$\,0.12   & 11.73\,$\pm$\,0.12 & 2015-04-11 & Bckg \\
TYC 9412-1370-1 & a & 1.801\,$\pm$\,0.003 & 61.57\,$\pm$\,0.16   & -- & 11.85\,$\pm$\,0.11   & 11.73\,$\pm$\,0.12 & 2016-04-03 & Bckg\\
HIP 69781 & a & 1.505\,$\pm$\,0.007 & 35.8\,$\pm$\,0.3   & 11.17\,$\pm$\,0.17 & -- & -- & 2013-04-26 & Bckg \\
HIP 69781 & b & 3.048\,$\pm$\,0.014 & 315.1\,$\pm$\,0.3   & 9.53\,$\pm$\,0.14 & --   & 11.73\,$\pm$\,0.12 & 2013-04-26 & Bckg \\
HIP 69781 & c & 3.956\,$\pm$\,0.018 & 293.3\,$\pm$\,0.3   & 10.68\,$\pm$\,0.12 & -- & -- & 2013-04-26 & Bckg \\
HIP 69781 & d & 4.62\,$\pm$\,0.02 & 166.3\,$\pm$\,0.3   & 11.91\,$\pm$\,0.14 & --   & -- & 2013-04-26 & Bckg \\
HIP 69781 & e & 4.78\,$\pm$\,0.02 & 55.0\,$\pm$\,0.3   & 12.78\,$\pm$\,0.12 & --   & -- & 2013-04-26 & Bckg \\
HIP 69781 & f & 5.20\,$\pm$\,0.03 & 140.6\,$\pm$\,0.3   & 13.30\,$\pm$\,0.14 & -- & -- & 2013-04-26 & Bckg \\
HIP 69781 & a & 1.5553\,$\pm$\,0.0014 & 34.73\,$\pm$\,0.14   & --  & 10.92\,$\pm$\,0.14   & 10.88\,$\pm$\,0.14 & 2016-04-12 & Bckg\\
HIP 69781 & b & 3.068\,$\pm$\,0.002 & 316.64\,$\pm$\,0.14   & -- & 9.23\,$\pm$\,0.14   & 9.18\,$\pm$\,0.14 & 2016-04-12 & Bckg\\
HIP 69781 & c & 3.950\,$\pm$\,0.003 & 294.44\,$\pm$\,0.14   & -- & 9.83\,$\pm$\,0.14   & 9.78\,$\pm$\,0.14 & 2016-04-12 & Bckg\\
HIP 69781 & d & 4.543\,$\pm$\,0.004 & 166.42\,$\pm$\,0.14   & -- & 11.87\,$\pm$\,0.14   & 11.82\,$\pm$\,0.14 & 2016-04-12 & Bckg\\
HIP 69781 & e & 4.781\,$\pm$\,0.005 & 54.98\,$\pm$\,0.14  & -- & 12.62\,$\pm$\,0.14   & 12.58\,$\pm$\,0.14 & 2016-04-12 & Bckg \\
HIP 69781 & f & 5.132\,$\pm$\,0.007 & 140.46\,$\pm$\,0.15   & -- & 13.12\,$\pm$\,0.14   & 13.07\,$\pm$\,0.14 & 2016-04-12 & Bckg \\
HIP 74049 & a & 5.54\,$\pm$\,0.03 & 148.3\,$\pm$\,0.3   & 12.20\,$\pm$\,0.14 & --   & -- & 2013-04-30 & Bckg \\
HIP 74049 & b & 7.31\,$\pm$\,0.04 & 221.9\,$\pm$\,0.3   & 14.09\,$\pm$\,0.14  & --   & -- & 2013-04-30 & Unknown\\
HIP 74049 & a & 5.425\,$\pm$\,0.004 & 145.82\,$\pm$\,0.14   & -- & 12.45\,$\pm$\,0.11  & 12.40\,$\pm$\,0.11 & 2015-04-11 & Bckg \\
1RXS J1535570 232417 & a & 3.395\,$\pm$\,0.003 & 73.13\,$\pm$\,0.14   & -- & 10.3\,$\pm$\,0.3  & 10.5\,$\pm$\,0.2 & 2017-03-05 & Unknown\\
1RXS J1535570 232417 & b & 4.228\,$\pm$\,0.004 & 318.48\,$\pm$\,0.14  & -- & 10.5\,$\pm$\,0.2  & 10.3\,$\pm$\,0.2 & 2017-03-05 & Unknown\\
HIP 77911 & a & 4.922\,$\pm$\,0.007 & 87.02\,$\pm$\,0.15  & -- & 14.1\,$\pm$\,0.9  & 14.0\,$\pm$\,0.8 & 2016-03-08 & Unknown\\
HIP 77911 & b & 4.230\,$\pm$\,0.003 & 228.38\,$\pm$\,0.14  & -- & 11.5\,$\pm$\,1.0  & 11.3\,$\pm$\,1.0 & 2016-03-08 & Unknown\\
HIP 77911 & c & 4.960\,$\pm$\,0.005 & 233.84\,$\pm$\,0.14  & -- & 13.2\,$\pm$\,0.9  & 12.9\,$\pm$\,0.8 & 2016-03-08 & Unknown\\
RXJ1601.9-2008 & a & 3.097\,$\pm$\,0.002 & 162.82\,$\pm$\,0.14  & -- & 11.33\,$\pm$\,0.11  & 11.13\,$\pm$\,0.11 & 2015-04-28 & Bckg\\
RXJ1601.9-2008 & b & 5.432\,$\pm$\,0.006 & 303.87\,$\pm$\,0.15  & -- & 13.28\,$\pm$\,0.11  & 13.08\,$\pm$\,0.11 & 2015-04-28 & Bckg\\
RXJ1601.9-2008 & a & 3.056\,$\pm$\,0.002 & 162.54\,$\pm$\,0.14  & -- & 11.07\,$\pm$\,0.11  & 11.05\,$\pm$\,0.11 & 2016-06-29 & Bckg\\
RXJ1601.9-2008 & b & 5.473\,$\pm$\,0.006 & 304.14\,$\pm$\,0.14  & -- & 13.01\,$\pm$\,0.11  & 13.12\,$\pm$\,0.11 & 2016-06-29 & Bckg\\
ScoPMS027 & a & 5.567\,$\pm$\,0.007 & 338.56\,$\pm$\,0.15  & -- & 12.0\,$\pm$\,0.2  & 12.0\,$\pm$\,0.2 & 2016-03-09 & Unknown\\
ScoPMS027 & a & 5.604\,$\pm$\,0.005 & 338.50\,$\pm$\,0.14  & -- & 11.6\,$\pm$\,0.2  & 11.7\,$\pm$\,0.2 & 2016-03-09 & Unknown\\
ScoPMS027 & b & 5.842\,$\pm$\,0.006 & 84.86\,$\pm$\,0.14  & -- & 11.5\,$\pm$\,0.2  & 11.5\,$\pm$\,0.2 & 2016-03-09 & Unknown\\
TYC 6209-735-1 & a & 6.312\,$\pm$\,0.005 & 322.59\,$\pm$\,0.14  & -- & 9.64\,$\pm$\,0.11  & 9.62\,$\pm$\,0.11 & 2016-03-09 & Unknown\\
TYC 6213-306-1 & a & 3.478\,$\pm$\, 0.002 & 76.50\,$\pm$\, 0.14  & -- & 11.50\,$\pm$\,0.10  & 11.36\,$\pm$\,0.11 & 2016-04-03 & Bckg\\
TYC 6213-306-1 & b & 3.883\,$\pm$\, 0.003 & 317.96\,$\pm$\, 0.14  & -- & 11.11\,$\pm$\,0.10  & 11.13\,$\pm$\,0.09 & 2016-04-03 & Bckg\\
HD 147808  & a & 2.215\,$\pm$\, 0.007 & 54.7\,$\pm$\, 0.2  & -- & 13.12\,$\pm$\, 0.11  & 13.17\,$\pm$\, 0.17 & 2016-04-14  & Unknown\\
HIP 84586  & a & 7.323\,$\pm$\, 0.007 & 109.26\,$\pm$\, 0.14  & -- & 11.87\,$\pm$\, 0.11  & 11.54\,$\pm$\, 0.11 & 2015-04-08  & Unknown\\
HIP 88481  & a & 5.155\,$\pm$\, 0.010 & 335.74\,$\pm$\, 0.17  & -- & 13.73\,$\pm$\, 0.11  & 13.73\,$\pm$\, 0.11 & 2015-05-17  & Bckg\\
HIP 88481  & a & 5.25\,$\pm$\,0.01 & 334.60\,$\pm$\, 0.17  & -- & 13.91\,$\pm$\, 0.11  & 13.94\,$\pm$\, 0.11 & 2016-04-14  & Bckg\\

HIP 94050  & a & 2.312\,$\pm$\,0.011 & 317.2\,$\pm$\, 0.3  & -- & 14.08\,$\pm$\, 0.14  & 13.97\,$\pm$\, 0.14 & 2015-04-22  & Bckg\\
HIP 94050  & a & 2.394\,$\pm$\,0.010 & 318.3\,$\pm$\,0.3  & -- & 14.22\,$\pm$\, 0.11  & 14.28\,$\pm$\, 0.11 & 2016-05-03  & Bckg\\

HIP 98704  & a & 3.208\,$\pm$\,0.002 & 138.80\,$\pm$\,0.1  & -- & 8.1\,$\pm$\, 0.2  & 8.1\,$\pm$\, 0.2 & 2015-05-28  & Bckg\\
HIP 98704  & b & 3.890\,$\pm$\,0.002 & 150.80\,$\pm$\,0.14  & -- & 8.6\,$\pm$\, 0.2  & 8.5\,$\pm$\, 0.2 & 2015-05-28  & Bckg\\
HIP 98704  & c & 5.233\,$\pm$\,0.004 & 66.65\,$\pm$\,0.14  & -- & 10.85\,$\pm$\, 0.13  & 10.77\,$\pm$\, 0.14 & 2015-05-28  & Bckg\\
HIP 98704  & a & 3.255\,$\pm$\,0.002 & 138.23\,$\pm$\,0.14  & -- & 9.29\,$\pm$\, 0.11  & 9.72\,$\pm$\, 0.11 & 2016-05-28  & Bckg\\
HIP 98704  & b & 3.927\,$\pm$\,0.003 & 150.19\,$\pm$\,0.14  & -- & 9.88\,$\pm$\, 0.11  & 9.72\,$\pm$\, 0.11 & 2016-05-28  & Bckg\\
HIP 98704  & c & 5.281\,$\pm$\,0.005 & 67.09\,$\pm$\,0.14  & -- & 12.14\,$\pm$\, 0.11  & 11.94\,$\pm$\, 0.11 & 2016-05-28  & Bckg\\

HIP 105404  & a & 3.645\,$\pm$\,0.016 & 37.8\,$\pm$\,0.4  & -- & 12.14\,$\pm$\, 0.11  & 11.94\,$\pm$\, 0.11 & 2013-06-28  & Bckg \\
HIP 105404  & b & 5.25\,$\pm$\,0.02 & 160.7\,$\pm$\,0.4  & -- & 12.14\,$\pm$\, 0.11  & 11.94\,$\pm$\, 0.11 & 2013-06-28  & Bckg\\
HIP 105404  & a & 3.707\,$\pm$\,0.003 & 34.60\,$\pm$\,0.14  & -- & --  & -- & 2015-05-28  & Bckg\\
HIP 105404  & b & 5.077\,$\pm$\,0.008 & 161.73\,$\pm$\,0.16  & -- & --  & -- & 2015-05-28  & Bckg \\

HIP 105860  & a & 4.937\,$\pm$\,0.012 & 168.03\,$\pm$\,0.15  & 13.93\,$\pm$\, 0.14 & --  & --  & 2011-10-10  & Bckg \\
HIP 105860  & b & 7.51\,$\pm$\,0.03 & 234.16\,$\pm$\,0.22  & 14.92\,$\pm$\, 0.14  & -- & -- & 2011-10-10  & Unknown\\
HIP 105860  & c & 8.262\,$\pm$\,0.019 & 225.46\,$\pm$\,0.15  & 14.18\,$\pm$\, 0.14 & --  & -- & 2011-10-10  & Unknown\\
HIP 105860  & a & 4.977\,$\pm$\,0.013 & 171.5\,$\pm$\,0.2  & -- & 14.05\,$\pm$\, 0.11  & 13.78\,$\pm$\, 0.11 & 2015-05-28  & Bckg\\

\hline
\end{tabular}
}
\end{table*}

\clearpage
\onecolumn

 \begin{center}
\longtab{
\footnotesize{
\begin{longtable}{l l c c c c c c }
\caption{\label{obslog} SPOTS Observing Log}\\
\hline\hline
HIP ID & Alt ID & Epoch & t$\rm_{tot}$(min)  & Rotation (deg)  & Cand Comp & Co-moving Comp & Instrument\\ \hline
\endfirsthead
\caption{continued.}\\
\hline\hline
HIP ID & Alt ID & Epoch & t$\rm_{tot}$(min)  & Rotation (deg)  & Cand Comp & Co-moving Comp & Instrument\\ \hline

\endhead
\hline\\
\endfoot

HIP 7601 \phantom{\Large!}  & HD 10800 &  2015-08-25 & 21.3 & 8.4 & 1 $\times$ star & ? & SPHERE\\
HIP 9892 & HD 13183& 2011-12-03&16.0 & 20.2 & -- & &NACO\\
 & & 2012-12-19&12.6 &  6.0 & -- & -- &NACO\\
HIP 12225 & $\eta$ Hor &  2015-08-27& 17.7& 17.9 &  1  & -- & SPHERE\\
 HIP 12545 & BD+05 378 &2011-10-31& 34.0 &27.1& -- &--&NACO\\
HIP 12716 & UX For & 2013-01-05 & 38.0 &43.9&1 $\times$ star& 1 $\times$ star&NACO\\
HIP 14007 & HD 18809 & 2015-08-29 & 25.6 & 25.2 & -- & -- & SPHERE \\
HIP 14568 & AE For &  2015-09-13&25.6 & 1.4 & -- & --& SPHERE\\
& & 2016-12-15 & 38.4 & 3.2 & -- & -- & SPHERE\\
HIP 14807 & BD+21 418B & 2015-09-08 & 25.6 & 8.4 & -- & -- & SPHERE\\
HIP 15197 & $\zeta$ Eri &   2015-08-29 & 17.1 & 22.2 & -- & -- & SPHERE\\
HIP 16853& HD 22705& 2011-11-09&26.6&18.4 & 1 $\times$ star & 1 $\times$ star & NACO\\
HIP 19591&V1136 Tau&2012-11-21& 26.0 & 8.8 & 1 $\times$ star& 1 $\times$ star&NACO\\
&TYC 5907-1244-1& 2011-11-05&20.4 & 67.1 & --&--&NACO\\ 
HIP 25486&AF Lep&2011-12-20&26.6&29.4& -- & -- & NACO\\
HIP 25709 && 2011-12-22&42&63.8& -- & --&NACO\\
HIP 27134&XZ Pic&2013-01-03&16&20.4 & -- &--&NACO\\
HIP 31681&Alhena&2011-12-22 & 22 & 11.7 & 2 &  & NACO\\
&&2013-03-01& 22.0 & 10.8 & 2 & -- & NACO\\
&TYC 8104-0991-1&2013-01-07 & 20.0 &11.4& 1 $\times$ star & 1 $\times$ star & NACO\\
HIP 32104&26 Gem&2012-01-13&27.3&10.0 & -- & & NACO\\
&&2013-01-05&42.3&17.6 & -- & -- & NACO\\
&EM Cha&2011-12-14&23.8&12.2& 1  & -- & NACO\\
&TYC-8569-3597-1&2011-12-17&20.0 &11.3 & CF &  &NACO\\
&&2015-04-11&34.1& 20.5 & CF &-- &SPHERE\\
&TYC 9399-2452-1&2013-03-03&20.4&13.9 & 1& & NACO\\
&&2016-04-03&52.3&7.5& 1 & -- & SPHERE\\
HIP 46509 & $\tau$ Hya & 2015-12-20 & 25.6&17.0 &--& --&SPHERE\\
	 & 	 & 2017-01-11 & 38.4 & 29.4 &--&--&SPHERE\\
HIP 46637&GS Leo&2013-02-10&26&10.5& -- & -- & NACO\\
HIP 47760&HD 84323& 2013-01-24&20.4&  27.4&-- &-- &NACO\\
&TYC 6604-0118-1&2013-01-26&20.4 &5.2& -- &-- & NACO\\
HIP 49669 & Regulus  & 2016-01-20 & 14.3 &12.0 & -- &  --& SPHERE\\
HIP 49809 & HD 88215 & 2015-04-25 &18.1 &26.3 &  -- & --& SPHERE\\
HIP 56960 & HD 101472 & 2015-04-11 & 21.3 & 24.0 & -- &--&SPHERE\\
HIP  57363 & $\lambda$ Muscae & 2016-01-24    & 25.6 & 10.8   & -- & & SPHERE\\
             &   & 2016-05-07   & 38.4 & 16.5    & 1 $\times$ star &  & SPHERE\\
          & & 2016-06-03   & 38.4  & 16.3    & 1 $\times$ star & 1 $\times$ star &SPHERE\\
              &TYC 9412-1370-1&2015-04-11&21.3&6.5 & 1 & & SPHERE\\
&&2016-04-03&34.1&10.6& 1& -- & SPHERE\\
HIP 63742 & PX Vir &  2013-02-11 & 48.7  & 41.4 & --  &  &NACO \\
 &   &  2015-04-11 &  25.6 &  21.6 & -- & -- & SPHERE\\
 HIP 69781 &V* V636 Cen &  2013-04-26  &27.3   &  6.9 & CF  &  &NACO \\
 & &  2016-04-12   & 25.6   &  15.0 &  CF  & -- &SPHERE \\
 HIP 74049&HS Lup&2013-04-30&26.6&17.6&2 & & NACO\\
&&2015-04-11&25.6&18.6&1 & 1?& SPHERE\\
&1RXS J153557.0-232417 & 2017-03-05 & 21.3 &  2.1 & 2 & ? & SPHERE\\
HIP 76629 & HD 139084&2012-07-20&27.8&14.3&CF&&NACO\\
&&2015-04-11&25.6&12.7&CF&--&SPHERE\\
HIP 77911 & HD 142315 & 2016-03-08 &25.6& 5.7 & 3 & ? & SPHERE\\
HIP 78416 & HD 143215 & 2013-05-11 & 12.0 &13.8& CF & & NACO\\
&&2015-04-22&12.8&13.8& CF&--&SPHERE\\
&RX J1601.9-2008& 2015-04-28& 25.6 & 69.2 & 2 & & SPHERE\\
&&2016-06-29 & 38.4 & 88.1 & 2 & -- & SPHERE\\
&ScoPMS027& 2016-03-09 &  21.3 & 32.0 & 1 & & SPHERE\\
&& 2016-04-07& 34.1&65.2& 2 & ? & SPHERE\\
&TYC 6209-735-1 &  2015-06-08 & 21.3 &45.9 & 1 & ? & SPHERE\\
&ScoPMS044& 2015-04-13&25.6&56.6 & -- & -- & SPHERE\\
&ScoPMS048& 2016-04-03&21.3&39.4 & 1 $\times$ star & 1 $\times$ star & SPHERE\\
&TYC 6213-306-1& 2016-04-03 &25.6& 3.8 & 2 & -- & SPHERE\\
&HD 147808 & 2016-04-14& 25.6 & 6.4 & 1 & ? & SPHERE\\
HIP 80686 & $\zeta$ TrA & 2015-04-08 & 21.9 & 10.6 & -- & -- & SPHERE\\
&ROX 33&2016-08-01&25.6&0.3 & -- & & SPHERE\\
&&2016-08-09&25.6 &0.8 & -- & -- & SPHERE\\
&ROXs 43A & 2016-08-27 & 38.4 & 1.1 & \small{1 $\times$ star + 1 $\times$ binary} & \small{1 $\times$ star + 1 $\times$ binary} & SPHERE\\
HIP 82747 & AK Sco & 2015-04-13 & 25.6 & 29.0 & CF + disk & & SPHERE\\
&&2016-04-01& 38.4 & 37.6& CF + disk & disk &SPHERE\\
HIP 84586 & HD 155555 &2015-04-08& 20.3& 10.8 & 1 & & SPHERE\\
&& 2016-04-08&29.9&15.6 & -- & ? & SPHERE\\
HIP 88481 & HD 165045 &  2015-05-17&25.6&15.1& 1 & & SPHERE\\
&&2016-04-14& 38.4 & 22.4 & 1 & -- &SPHERE\\
HIP 94050 & HD 177996 &2015-04-22 &25.6&17.9 & 1 & & SPHERE\\
 & &2016-05-03 &38.4&32.3 & 1 &-- & SPHERE\\
HIP 98704 & HD 188480 & 2015-05-28 & 25.6& 7.9 & 3 &&SPHERE\\
 &  & 2016-08-04 & 38.4& 12.3 & 3 &--&SPHERE\\
HIP 100751 & $\alpha$ Pav&2015-07-18&21.3&10.1& -- & -- & SPHERE\\
HIP 101422 &HD 195289&2015-05-17& 25.6&12.3& --& --&SPHERE\\
HIP 104043 & $\alpha$ Oct &2015-06-08&12.8&8.8&--&--&SPHERE\\
HIP 105404&BS Ind & 2013-06-28& 26&14.1 & 2 &  &NACO\\
& & 2015-05-28&  4.26 &  0.7 &  2& --& SPHERE\\
HIP 105860 & IK Peg & 2011-10-10 & 32.3 & 16.7 & 3 & & NACO\\
 &&2015-05-28 &25.6 &6.9 & 1 & -- & SPHERE\\
HIP 107556&$\delta$ Cap & 2011-10-06 & 22.0 & 50.6 & -- &--& NACO\\
HIP 109110 & NT Aqr &  2015-06-11 & 25.6 & 19.5 & -- &--&SPHERE\\
HIP 109901 & CS Gru & 2013-06-26& 12.0 & 11.4 &--&--&NACO\\
&TYC 6386-0896-1 & 2012-11-26&20.4 & 6.1 & -- & --&NACO\\
HIP 113860 & pi. PsA & 2015-06-20 & 10.3 &  39.5 & -- &--&SPHERE\\

\hline
\end{longtable}}
\tablefoot{
Resolved companion candidates to each observed tight binary. For each individual observation we present the epoch, total exposure time, accumulated sky rotation, the presence of candidate companions and the corresponding
instrument used to acquire it. CF = crowded field}
}
\end{center}

\end{appendix}

\end{document}